\documentclass[manuscript,twocolumn]{aastex63}
\usepackage{float}
\usepackage[T1]{fontenc}   
\usepackage{amsmath}  
\usepackage{graphicx}  
\usepackage{xcolor}

\revised{\today}

\begin{document}
\shorttitle{ALMA observations of Carina's Western Wall} \shortauthors{Hartigan et al.}

\title{ALMA Datacubes and Continuum Maps of the Irradiated Western Wall in Carina
\vspace{0.5 in}}

\correspondingauthor{Patrick Hartigan} \email{hartigan@sparky.rice.edu}

\author[0000-0002-5380-549X]{Patrick Hartigan} \affiliation{Physics and Astronomy Dept., Rice University, 6100 S. Main, Houston,
TX 77005-1892}

\author{Maxwell Hummel}

\author[0000-0001-8061-2207]{Andrea Isella} \affiliation{Physics and Astronomy Dept., Rice University, 6100 S. Main, Houston, TX
77005-1892}

\author[0000-0002-7639-5446]{Turlough Downes} \affiliation{Centre for Astrophysics \& Relativity, School of Mathematical Sciences,
Dublin City University, Glasnevin, Dublin 9, Ireland}

\begin{abstract} 
We present ALMA observations of the continuum and line emission of
$^{12}$CO, $^{13}$CO, C$^{18}$O, and [C~I] for a portion 
of the G287.38-0.62 (Car 1-E) region in the Carina star-forming 
complex.  The new data record how a molecular cloud
responds on subarcsecond scales when subjected to a 
powerful radiation front, and provide insights into
the overall process of star formation
within regions that contain the most massive young stars.
The maps show several molecular clouds superpose upon
the line of sight, including a portion of the Western Wall,
a highly-irradiated cloud situated near
the young star cluster Trumpler 14.  In agreement with             
theory, there is a clear progression from fluoresced
H$_2$, to [C~I], to C$^{18}$O with distance into the PDR front.
Emission from optically thick $^{12}$CO extends across
the region, while $^{13}$CO, [C~I] and especially C$^{18}$O are more optically thin, 
and concentrate into clumps and filaments closer to the PDR interface.
Within the Western Wall cloud itself we identify 254
distinct core-sized clumps in our datacube of C$^{18}$O. The
mass distribution of these objects is similar to that of the stellar IMF.
Aside from a large-scale velocity gradient, the clump radial velocities  
lack any spatial coherence size. There is no
direct evidence for triggering of star formation
in the Western Wall in that its C$^{18}$O clumps and continuum cores
appear starless, with no pillars present. However,
the densest portion of the cloud lies closest to the PDR,
and the C$^{18}$O emission is flattened along the radiation front.

\end{abstract}

\keywords{Photodissociation regions (1223), Star formation (1569), radiative magnetohydrodynamics (2009)}

\section{Introduction} 
\label{sec:intro}

The first surveys of molecular clouds established that the densest
portions of the clouds, known as cores, mark locations where
stars are born \citep[e.g.][]{Myers83}. These observations make sense
from a theoretical standpoint, as the densest parts of the cloud
should be the first areas to collapse gravitationally, and
numerical models predict that within quiescent regions core collapse should
begin within its inner regions and then expand outward \citep[e.g.][]{Shu77,Shu87}. As
material accumulates, the central density and temperature of the core 
should rise until a protostar forms, and the object subsequently evolves along an
evolutionary track in the HR-diagram determined primarily by its mass.

The decades since these early studies have witnessed an avalanche of new data 
related to star formation as space-based missions such as IRAS,
Spitzer, Herschel, and WISE have surveyed the galactic plane
across the entire infrared and far-infrared spectral bandpasses,
while targeted ground-based studies in both continuum and various
molecular emission lines provided deep maps with high spatial
resolution of the dust distribution, molecular content, and dynamics within
individual star-forming regions.
Although the general scenario that stars form in molecular cloud cores
remains intact, the new observations have made it
clear that the actual process of star formation can be incredibly complex.
In many and perhaps most regions, the factors behind this complexity
seem to play a dominant role in determining
whether or not a star forms at all, and 
also influence the masses, compositions, and orbital radii of 
planets that coalesce out of the protostellar disk material remaining after
the star forms.

One of the main contributing factors to the complexity in star formation regions
is simply the lack of spherical symmetry within molecular clouds.
Recent parsec-scale maps with Herschel have shown that dust in molecular
clouds appears highly filamentary \citep{Hill11,arzou11}, and velocity maps
along these filaments display large-scale organized motions \citep{Peretto14,Kong19},
clearly demonstrating that simple 1-dimensional
models of collapse do not represent the physics accurately.
Polarization maps sometimes indicate ordered magnetic fields on large scales that 
could affect the dynamics, while other regions 
appear more chaotic and turbulent \citep{planck16,Hull20}.
High-resolution maps of molecular clouds now available with ALMA
also show geometrical complexity on small-scales \citep[$\lesssim$ 0.01 pc][]{Tokuda20},
where ambipolar diffusion is important to consider when assessing magnetic pressure support
for collapsing cores \citep{Chen14}. The additional physics introduced by these
processes affects the efficiency of star formation within a cloud as
well as the time scale for collapse.
Geometrical factors are particularly important for massive stars, as
cavities created along the rotation axes provide a means for
radiation to escape while accretion is ongoing through a disk, without which it
would be impossible to accumulate enough material
to create the highest mass stars \citep{Rosen16,Krumholz05}.

Once stars form, the situation becomes even more complex because 
radiation from the newborn stars and collimated bipolar
outflows deposit energy back into the surrounding cloud.
This energy could cause the gas in the nascent cloud
to disperse or drive turbulence into the cloud
that adds a pressure term to counter gravitational collapse, effectively lowering the star
formation rate. On the other hand, shock waves from outflows and radiation fronts also 
cause density enhancements, which in turn might trigger stars to form by pushing
pre-existing clumps past their Jeans limit \cite[e.g.][]{Haworth11}.
On large scales, because there is
a finite amount of gas within the molecular cloud, a core that might otherwise 
form one or more stars may become starved of gas if a nearby area collapses first
\citep{bonnell03}.
These competitive accretion scenarios appear in numerical simulations, and
are an attractive way to explain the morphologies of
some regions \citep{Wang10,Sanhueza19}. Such large-scale environmental factors set the initial conditions
for any core collapse, after which the atomic physics of molecular gas, dust, and ionization
determine how rapidly the material cools and the degree to which ambipolar diffusion 
operates, which in turn influence fragmentation scales and ultimately the mass
function of newborn stars and protostellar disks that emerge from the molecular cloud.
Some recent theoretical studies indicate that the most massive stars 
originate from ridges in a molecular cloud that
develop asymmetrical gas inflow streams \citep{motte18}.

Our understanding of cores has increased dramatically in the past few years owing to
the availability of wide-field, high-resolution surveys of nearby star-forming regions 
such as Perseus \citep{EnochPer,PerseusCores}, Aquila \citep{AquilaCores},
Chameleon \citep{ChameleonCores}, Corona Australis \citep{AustralisCores}, Lupus
\citep{LupusCores}, Taurus \citep{TaurusCores}, Monoceros \citep{MonocerosCores},
and Orion A \citep{OrionACores} that identify cores as cold ($\sim$ 20~K)
continuum sources.  Observations of molecular lines complement core studies by
providing intensity and velocity maps in the form of data cubes, which yield
insights into turbulent size scales and internal motions \citep[e.g.][]{Roman11,Chen19},
and fine-structure lines in atomic transitions such as
[C~I] 609 $\mu$m probe regions of atomic gas that may surround
or intermix with molecular gas \citep{Tomassetti14}.
Mass distributions of cores in low-mass star formation regions like those above
generally follow the overall stellar initial mass function, suggesting
a constant conversion efficiency from core mass to stellar mass \citep{Andre17,Orion20}.
However, even Orion~A does not sample environments where the most
massive early-O stars form and the effects of radiation will be strongest.
We must begin to study cores in these environments if we are to 
understand the role radiation plays in forming stars.

Anchored by the extremely massive luminous blue variable
star $\eta$ Car and home to more than 100 O-stars and early B-stars
\citep{Walborn02,Smith06} and ongoing star formation 
\citep{broos11,preibisch11a,smith10,gaczkowski13},
Carina OB1 is an ideal star-forming environment for studying radiative processes.
Visible to the unaided eye, the Carina Nebula extends for almost two degrees
on the sky \citep[$\sim$ 80~pc for a distance of 2.3~kpc][]{Lim19},
and contains a great deal of structure. Numerous surveys of the clouds
in Carina, along with their associated catalogs and nomenclature,
exist across the electromagnetic spectrum.
The first low-resolution radio continuum surveys showed two main peaks, one
associated with $\eta$ Carina (Car~II) and another with an
H~II region to the west of the Tr~14 cluster \citep[Car~I ; ][]{gardner70},
which later observations by \citet{whiteoak94} resolved into three peaks,
Car~I-E, Car~I-W, and Car~I-S, atop a broad plateau. 
Car~I-E (G287.38-0.62) is an H~II region associated with a bright-rimmed
dark cloud that extrudes from
Carina's main dark lane in the optical \citep{deharveng75}.
\citet{brooks03} surveyed Car~I in CO with SEST 
($\sim$ 20\arcsec\ resolution), and catalogued several
clouds with angular sizes several beam widths in extent. This work also mapped 
[O~I] and [C~II] emission lines associated with photodissociation region
(PDR) of the main dark cloud
in Car~I-E using the KAO, and similar hydrogen recombination line maps 
also exist for the Car~I-E H~II region
\citep{brooks01}.  A mm-continuum survey of the Carina Nebula for clumps at 870$\mu$m 
with an 18\arcsec\ beam with the 12-m APEX dish found a power law spectrum with index $-1.95$,
similar to that expected for the initial mass spectrum for high-mass stars \citep{pekruhl13},
and a large-scale map with the same instrument showed the warm dust associated
with the irradiated cloud \citep{preibisch11b}.
Carina was also included in the MOPRA survey, which had a 40\arcsec\ beam and
observed dense tracers such as C$^{18}$O and HCO+ \citep{barnes11}.

Irradiated interfaces particularly stand out in near-IR H$_2$ images because
FUV photons are absorbed into the Lyman and Werner bands of H$_2$
within the PDR, and the subsequent fluorescence emits 2.12 $\mu$m photons
which can penetrate through the intervening dust. Continuum-subtracted H$_2$
images have $\sim$ 20 times the spatial resolution of pre-ALMA molecular maps,
and uncovered several walls and fat pillars situated to the east, north,
southwest, northwest and west of $\eta$ Car and Tr~14 
\citep[regions 9-15, 16-22, 44-50, 51-59, and 60, respectively of ][]{hartigan15}.
\citet{menon21} observed several of these pillars recently with ALMA, and 
found evidence for compressive turbulence induced by the
radiation fronts.
The brightest irradiated interface in the region as defined by the H$_2$ emission,
aptly named the `Western Wall', lies to the west of Tr~14.
This object is the bright-rimmed dark cloud adjacent to the Car~I-E H~II region noted above,
and is the source of the PDR emission lines that have been observed at this
location at low spatial resolution. 

As we will show in this paper,
high-resolution ALMA maps resolve several molecular clouds with differing
radial velocities along the line of sight to the Car~I-E region. To avoid
confusion, we use the term `Western Wall' 
to refer to only the irradiated cloud in Car~I-E.
Despite appearing rather indistinct in the optical owing to foreground dust
extinction, the Western Wall is one of the brightest objects in Carina when
observed in the fluorescent lines of H$_2$ and recombination lines of
Br$\gamma$ \citep{hartigan15,hartigan20}. It is a large structure about 2 arcminutes
($\sim$ 1.3~pc) in extent, illuminated by highly luminous, massive stars such as the O2 star 
HD~93129 in the nearby cluster Trumpler~14 \citep{HD93129,brooks03}, and 
$\eta$ Car in the Trumpler 16 cluster \citep{Wu18}. The cloud emits in a relatively
narrow velocity range of a few km$\,$s$^{-1}$, making it possible to isolate
the Western Wall cloud from the other clouds in datacube observations of the Car~I-E region.

This paper presents new mm-continuum and emission-line maps acquired with ALMA
in the region of southern portion of Carina's Western Wall.  Our goal is to 
quantify how a strong radiation field influences the development of star formation.
With ALMA's unprecedented resolution we can look for
spatial offsets of emission lines predicted in this photodissociation region,
explore the dynamics within the cloud at various optical depths, and
resolve features as small as 0.01 pc. This combination gives
us an exciting look at the effects that ionizing radiation
has on molecular clouds on both large and small scales.

The paper is organized as follows.
In Section 2 we discuss the acquisition and processing of the ALMA emission line
and continuum observations at 0.62 mm and 1.33 mm,
and Section 3 presents an overview of the continuum maps and data cubes we acquired in 
$^{12}$CO, $^{13}$CO, C$^{18}$O, and C~I.
This section derives maps of the optical depths within the region, considers spatial
offsets along the PDR, estimates a mass for the cloud, and examines continuum maps to
constrain sizes of the dust grains.
Section 4 focuses on identifying and characterizing clumps and cores within the region.
Section 5 assesses the effect that radiation may have on star formation in
the irradiated cloud, and considers whether or not protostars have formed in the
area of our maps.  The final section brings together
the main conclusions for the paper.

\section{Acquisition and Processing of ALMA Data}
\label{sec:data}

\begin{figure*}[!t]
\centering
\includegraphics[width=1.0\textwidth]{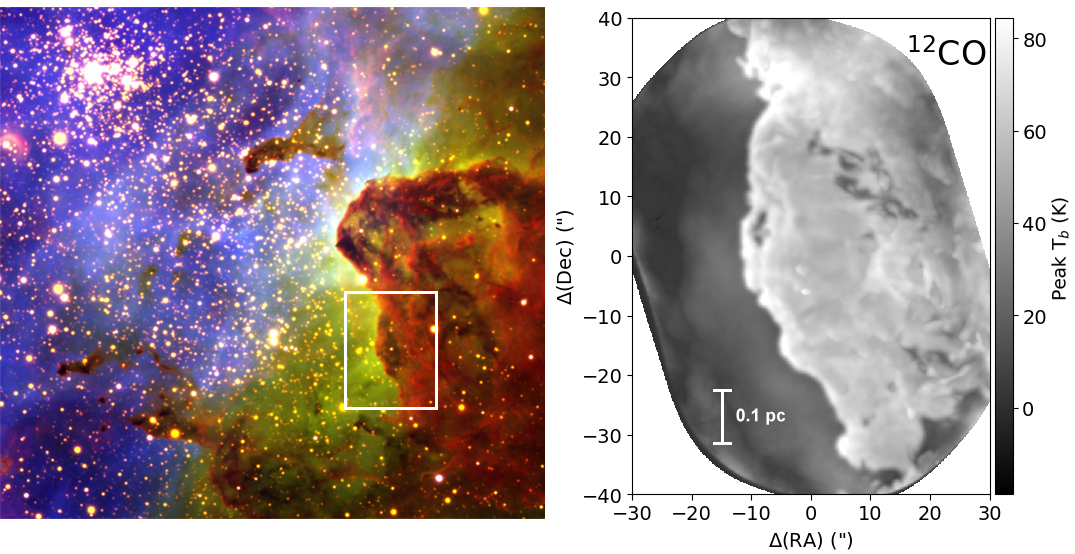}
\caption{Left: Composite RGB image of the Western Wall region in the
Carina nebula from \citet{hartigan15}. The image spans approximately
4.25~pc in RA and 4.0~pc in DEC. Red, green, and
blue colors indicate emission from H$_2$, Br$\gamma$, and [O III], respectively.
The stellar cluster at the top-left is Trumpler 14. The white rectangle outlines
the region expanded at right. Right: Peak brightness temperature in degrees K 
in the velocity interval $-$28.5 km$\,$s$^{-1}$ to $-$19.0 km$\,$s$^{-1}$ in 
$^{12}$CO J = 2 - 1 recorded by ALMA toward the Western Wall.
The angular resolution of the ALMA observations (FWHM of the synthesized beam)
is 1.0\arcsec\ for $^{12}$CO J = 2 - 1.
The spatial coordinates $\Delta$RA and $\Delta$DEC are relative to the
central map position of $\alpha$(2000) = 10:43:30.64,
$\delta$(2000) = $-$59:35:57.4.}
\label{fig:composite}
\end{figure*}

We used the Atacama Large Millimeter / Submillimeter Array (ALMA) on several dates
between December 2015 and September 2016 to map a portion of Carina's Western Wall.
The maps, centered at $\alpha$(2000) = 10h43m30.64s, $\delta$(2000) = $-$59\arcdeg 35\arcmin 57.4\arcsec\
(see Fig.~\ref{fig:composite}), covered a region $\sim$ 65\arcsec$\times$80\arcsec\
in size, or 0.7~pc $\times$ 0.9~pc at a distance of 2.3 kpc \citep{Lim19}. 
ALMA's receivers were tuned to a wavelength of 1.33~mm (ALMA Band~6) to
record the emission from the J = 2 - 1 transitions of $^{12}$CO, $^{13}$CO, and C$^{18}$O,
and to 0.62~mm (ALMA Band 8) for [C~I] 609 $\mu$m. 
In Band 6, we observed the J = 2 - 1 transitions of
$^{12}$CO 230.5380000 GHz ($\lambda$ = 1.30030~mm),
$^{13}$CO 220.3986841~GHz ($\lambda$ = 1.36017~mm),
and C$^{18}$O 219.5603541~GHz \citep[$\lambda$ = 1.36637~mm;][]{lamda}.
Each line was recorded with 0.122~MHz wide channels, corresponding
to a velocity resolution of 0.166 km s$^{-1}$.

In Band 8, observations at
492.160651~GHz \citep[$\lambda$ = 609.15 $\mu$m;][]{Harris17} and
489.750921~GHz \citep[$\lambda$ = 612.15 $\mu$m;][]{Gottlieb03} 
provided measurements of the [C~I] ($^3$P$_1-^3$P$_0$)
and CS J = 10 - 9 lines, respectively.
Both lines were recorded using 0.488 MHz wide channels, corresponding to
a velocity resolution of 0.297 km$\,$s$^{-1}$, and the datacubes rebinned to
0.30 km$\,$s$^{-1}$.
Our data set includes continuum settings in both bands. At Band 6,
the ALMA correlator was set up with two 1.875~GHz-wide
continuum bands centered at 231.602~GHz and 218.482~GHz, respectively, while in Band 8, 
two 1.875~GHz wide bands centered at 478 GHz and 480 GHz gave continuum.

The observations utilized the 12-m, 7-m, and total power antennas.
For Band 6, the 12-m array achieved maximum baselines of 390~m, 460~m, and 6.3~km in three
separate epochs, and the 7-m data derives from a compact configuration with maximum
baselines of about 45~m.  This combination of 12-m, 7-m, and total power data probes
emission on spatial scales as small as about 0.04\arcsec\ at 1.3~mm. 
We acquired Band 8 data in a single epoch when the 12-m array was on an
extended configuration with baselines up to 6.3~km, resulting in a
theoretical maximum angular resolution of 0.02\arcsec.  Small mosaics
of 5 and 13 pointings for the 7-m and 12-m arrays at Band 6, and 22 and 57
pointings for the 7-m and 12-m arrays at Band 8 sufficed to cover the region of interest.
In all cases, mosaic pointings were spaced by 0.511$\times$HPBW
(Half Primary Beam Width) to sample the emission at the Nyquist spatial
frequency.  We calibrated the data using version 4.7.0 of the ALMA pipeline except for the
Band 8 data from the 12-m array, which we acquired as a non-standard mode
and calibrated manually using CASA 4.7.2.  Table~\ref{tab:obs} in the Appendix
lists the calibrators we used to derive complex phase and amplitude gains, and 
also presents the observing log of the entire data set.

We combined the single-dish and interferometric observations
for the line emission in the Fourier space using a modified version of TP2VIS \citep{tp2vis}
as described in the Appendix.  Images of the continuum emission used only 7-m and
12-m array interferometric data.  The task TCLEAN (CASA Version 5.6.0) generated
both the continuum and line images.  We performed image deconvolution in Band 6
using multi-scale cleaning with Briggs weighting and a robust parameter equal 1.
This procedure resulted in a synthesized beam size of 0.7\arcsec\ in the continuum
and 1\arcsec\ in the CO lines.  For band 8, the continuum reductions adopted a
Briggs weighting with robust 0.3 and applied a uv-taper corresponding to an
on-sky FWHM of 0.25\arcsec. UV-tapering the band 8 data is necessary to remove
the longest baselines and to achieve a synthesized beam with low side lobes.
The angular resolution of the final Band 8 images is 1.2\arcsec$\times$0.9\arcsec\
in the line emission and 1.1\arcsec$\times$1.4\arcsec\ in the continuum.  

\section{Emission Line and Continuum Maps of $^{12}$CO, $^{13}$CO, C$^{18}$O and [C I]}
\label{sec:emission-maps}

In this section we present results of the emission-line and continuum maps
of the area outlined in Fig.~\ref{fig:composite}.
Sec.~\ref{sec:obs_line} identifies individual clouds superposed along the line of sight
from their radial velocities and considers several compact sources visible 
in the $^{12}$CO datacube. Sec.~\ref{sec:ww} calculates [C~I] and CO optical depths
and abundance ratios, compares the [C~I] data with CO, estimates a mass for the Western Wall cloud,
examines spatial offsets between the emitting layers of the PDR, and 
provides an overview of the kinematics within the Western Wall cloud. 
Sec.~\ref{sec:cont} considers what the continuum observations of the region reveal 
about grain size, and investigates how the PDR influences the morphology of
the densest concentration of clumps and cores in the area.

\subsection{Overview of the Clouds and Compact Sources in the Map}
\label{sec:obs_line}

%

\begin{figure*}[ht]
\centering
\includegraphics[width=0.4\linewidth]{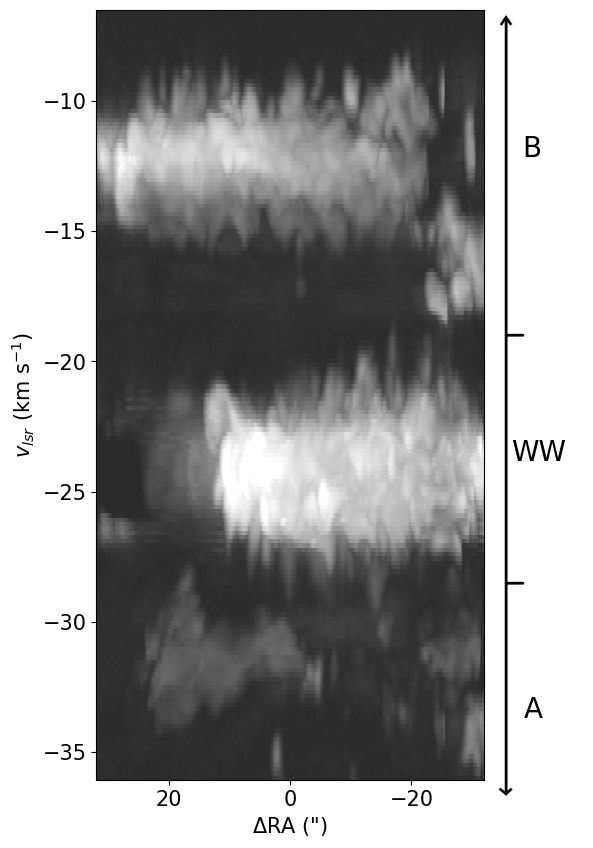}
\includegraphics[width=0.58\linewidth]{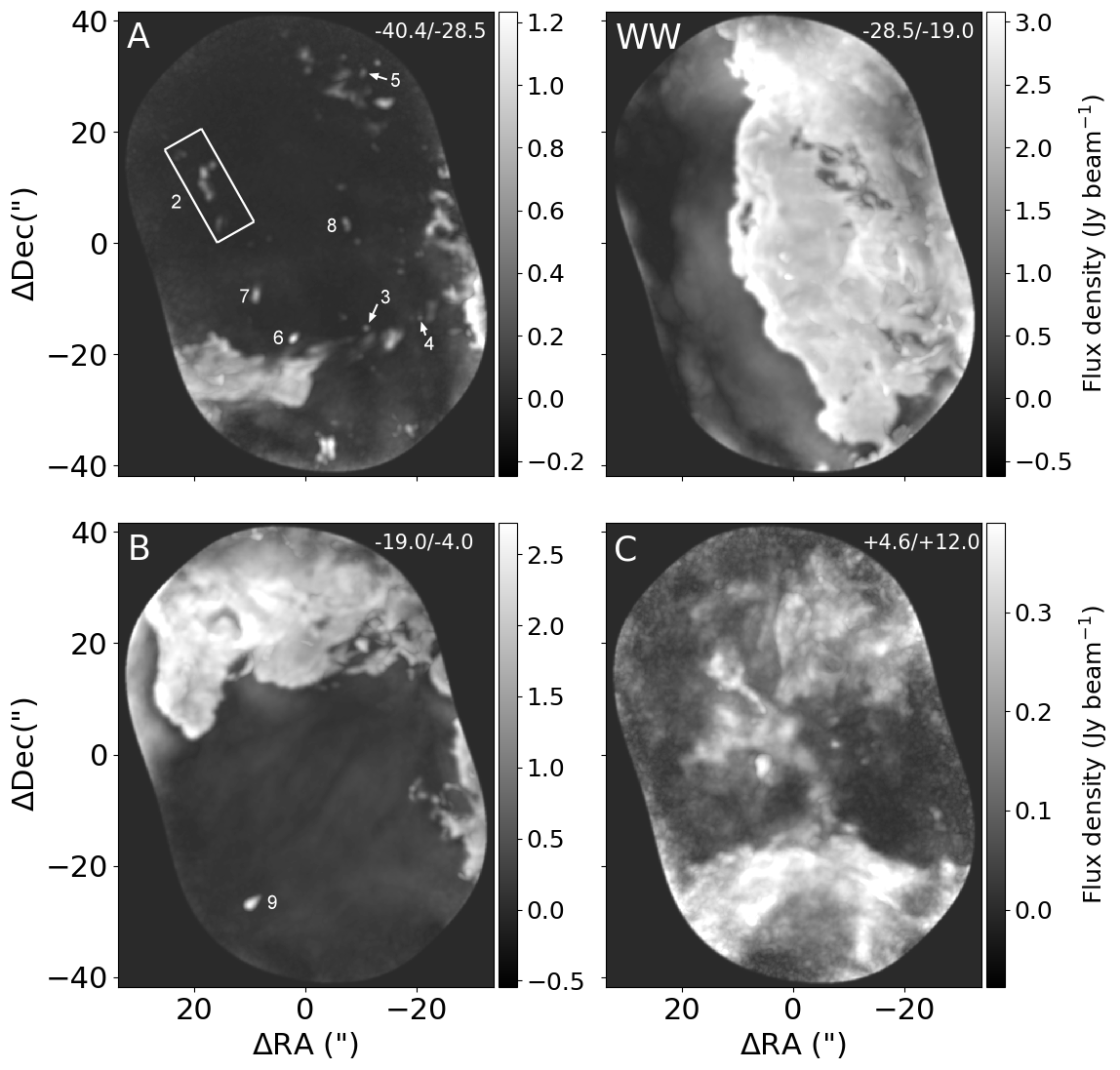}
\caption{Left: position-velocity diagram of $^{12}$CO 
recorded in the area shown in Fig.~\ref{fig:composite}, integrated
over all declinations in the map.
Right: maps of the peak intensity 
in the velocity range indicated in the top right of each panel.  The
$^{12}$CO emission separates into four distinct ranges in velocity.
Velocity range A includes
emission from $-$40.5 km$\,$s$^{-1}$ to $-$28.5 km$\,$s$^{-1}$, and contains
several compact sources labeled with numbers
as well as two clouds that appear unrelated to the Western Wall. 
The Western Wall cloud emits from $-$28.5 km$\,$s$^{-1}$ to $-$19.0 km$\,$s$^{-1}$ in CO, 
with most of the emission in the range from 
$-$21.3 km$\,$s$^{-1}$ to $-$27.1 km$\,$s$^{-1}$,
and extends from north to south across the field of view.
Region B, spanning $-$19.0 km$\,$s$^{-1}$ to $-$4 km$\,$s$^{-1}$, shows molecular
clouds that reside within Carina OB1 but are unrelated to the photodissociation front. 
Clouds with velocities between
+4.6 and +12.0 km s$^{-1}$ (region C), are diffuse and cover the entire field. 
The spatial offsets are defined relative to the
central map position in Fig.~\ref{fig:composite}.}
\label{fig:pos-vel}
\end{figure*}

\begin{figure*}[t]
\centering
\begin{interactive}{animation}{12CO_movie_3.mp4}
\includegraphics[width=0.5\linewidth]{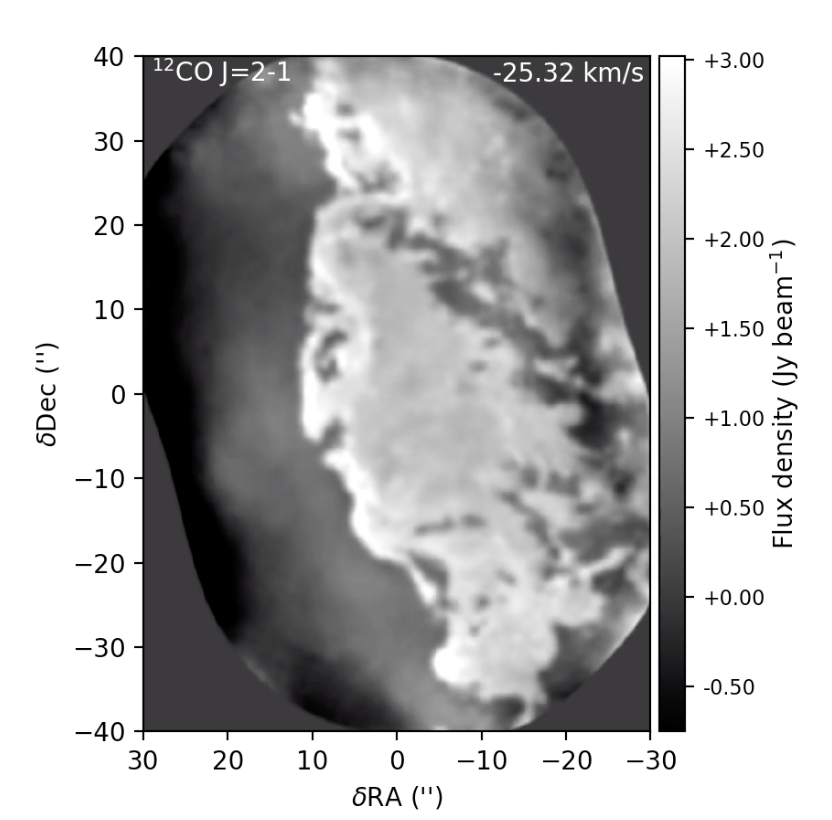}
\end{interactive}
\caption{Animation through the channels in the ALMA $^{12}$CO datacube.
The image is centered at $\alpha$(2000) = 10h43m30.64s,
$\delta$(2000) = $-$59\arcdeg 35\arcmin 57.4\arcsec, and ranges from
$-$45.16 km$\,$s$^{-1}$ to $+$5.16 km$\,$s$^{-1}$ V$_{lsr}$
in increments of 0.166 km$\,$s$^{-1}$. Each frame is normalized to the
value indicated by the scale bar at right.  The video duration is 75 seconds.
}
\label{fig:movie}
\end{figure*}

Fig.~\ref{fig:composite} shows that the region mapped by our ALMA observations covers
roughly the southern half of the Western Wall, easily visible in the
optical/infrared composite as a bright irradiated cloud characterized by 
H$_2$ fluorescence at the PDR interface and Br$\gamma$ emission that arises in the
photoevaporative flow \citep{hartigan15,hartigan20}. 
A compact cluster of young stars known as Trumpler 14 situated
about 3.5\arcmin\ to the northeast of the Western Wall is the main source of irradiation,
though additional massive stars, including $\eta$ Car and others in the young cluster Trumpler 16 located
off the frame to the southeast, also irradiate the cloud.

The position-velocity diagram of $^{12}$CO emission shown in Fig.~\ref{fig:pos-vel} and in the
animation in Fig.~\ref{fig:movie} uncovered several distinct molecular clouds that
lie within the area mapped by our observations. We refer to these features as `clouds'
rather than `clumps' because their radial velocities differ markedly and they are
unlikely to be associated with one-another. We use the term `clump' to refer
to small structures within our datacubes (see section~\ref{sec:c18oclumps}). 
Carina is located within a degree of the galactic plane, so it would not be unusual to
observe multiple molecular clouds projected along the line of sight.

Beginning with the most negative
velocities in the data cube (panel A in Fig.~\ref{fig:pos-vel}), the first extended
emission from molecular clouds in the region appears around $-$35 km$\,$s$^{-1}$ at the far southwestern
edge of the mapped region near ($\Delta$RA,$\Delta$Dec) = ($-$25$^{\prime\prime}$,$-$15$^{\prime\prime}$). Because
this feature only occurs at the edge of our map and
is superposed spatially upon the molecular cloud that defines the Western Wall, it is
difficult to identify any counterpart in the optical/IR composite with certainty.
A second coherent structure in Panel A spans $\sim$ $\Delta$RA = 0$^{\prime\prime}$ to +15$^{\prime\prime}$,
$\Delta$DEC = $-$30$^{\prime\prime}$ to $-$20$^{\prime\prime}$ and velocity
range $-$29 km$\,$s$^{-1}$ to $-$34 km$\,$s$^{-1}$. This feature may be associated
with a small area of dust that appears to lie in front of the H~II region in
Fig.~\ref{fig:composite}. In any event, neither of these objects seems to be
associated with the irradiated wall.

The dark cloud that makes up the irradiated Western Wall ranges in velocity between
about $-$19 km$\,$s$^{-1}$ and $-$28.5 km$\,$s$^{-1}$.
At the edges of its velocity range, the Western Wall molecular cloud breaks
into fragments, probably
because only the densest portions are optically thick at those velocities. 
The Western Wall becomes relatively
smooth around $-$24 km$\,$s$^{-1}$ where the entire wall is optically thick.
Overall, the shape of the Western Wall traced in CO follows the shape of
the irradiated interface defined by the fluoresced H$_2$ emission.
We discuss the velocity field within the Western Wall more fully in the next section.

The molecular cloud visible in panel B of Fig.~\ref{fig:pos-vel} 
($-$19 km$\,$s$^{-1}$ to $-$4 km$\,$s$^{-1}$) along the northern edge of
the map (hereafter `Cloud B') is associated with the Carina complex, as
the edge of the cloud appears weakly in the H$_2$ and Br$\gamma$ images. 
However, this cloud is located behind the Western Wall, based on the extra Br$\gamma$ emission
in this area as compared to what occurs at the position of the Western Wall molecular
cloud, and the fact that the Western Wall blocks out what we can see of Cloud
B in Fig~\ref{fig:composite}. Cloud~B provides a good comparison for the Western Wall cloud,
in that both are located in the Carina complex but the two clouds
have markedly differing radiation environments. In this velocity range there
is also a strip of CO emission that runs
vertically along the far western edge of the map and corresponds to
the edge of a feature that also appears weakly in archival Spitzer images of
the region \citep{tapia15,brooks03}. This emission arises from
another, probably foreground molecular cloud in this area.

\begin{figure*}[!t]
\centering
\includegraphics[width=1.0\linewidth]{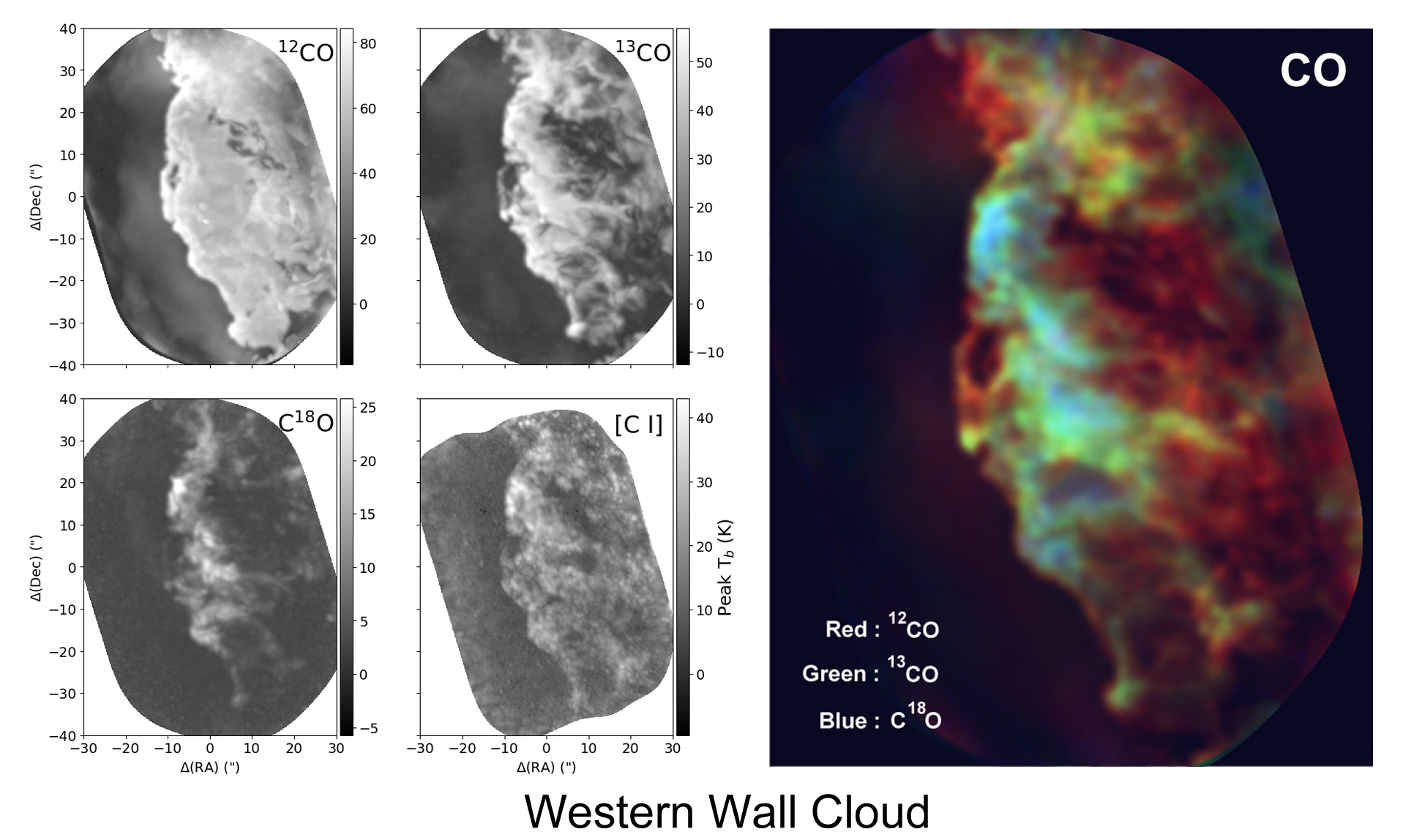}
\caption{
Left: Maps of the peak intensity of the $^{12}$CO, $^{13}$CO, C$^{18}$O, and [C~I] lines expressed in
units of the brightness temperature (K).  Right: Color composite of the CO emission from the
Western Wall molecular cloud, constructed by integrating the datacubes of
$^{12}$CO (red), $^{13}$CO (green), and C$^{18}$O (blue)
between V$_{lsr}$ velocities $-$21.28 km$\,$s$^{-1}$ and $-$27.09 km$\,$s$^{-1}$.
The densest portions of the cloud (areas with the strongest
C$^{18}$O emission) occur closer to the PDR interface relative to
the $^{12}$CO image, which is optically thick everwhere.
}
\label{fig:co_isotope}
\end{figure*}

At velocities more positive than those of the clouds in panel B of Fig.~\ref{fig:pos-vel}, the 
data cube is empty for a span of nearly 10 km$\,$s$^{-1}$ until an array of
molecular clouds begins to fill the field (panel C). These clouds have no counterparts
in the optical/IR composite, and are most likely to be background objects.

\begin{deluxetable*}{cccccl}[!ht]
\tablecaption{Compact Sources in the $^{12}$CO Datacube\label{tab:compact}}
\tablehead{
\colhead{Source}
 & \colhead{$\alpha$ (2000)}\vspace {-0.08in}
 & \colhead{$\delta$ (2000)}
 & \colhead{V$_{{\rm LSR}}$ (km$\,$s$^{-1}$)}
 & \colhead{IR Counterpart}
 & \colhead{Notes}\\
 (1)\vspace {-0.1 in}&(2)&(3)&(4)&(5)&\hbox to 1.45in{\null\hfil(6)}\\
}
\startdata
\noalign{\smallskip}
1&10:43:28.84& $-$59:36:12.0& $-$42.0& H$_2$ nebula & \\
2&10:43:33.09& $-$59:35:44.4& $-$37 to $-40$& none& An arc with multiple knots; N-S velocity gradient\\
3&10:43:29.22& $-$59:36:12.6& $-$39.1& dark globule& \\
4&10:43:27.94& $-$59:36:10.8& $-$38.4& dark globule& \\
5&10:43:29.29& $-$59:35:26.5& $-$36.4& none& \\
6&10:43:30.93& $-$59:36:14.3& $-$36.2& dark globule& Bipolar, PA $\sim$ 315 degrees\\
7&10:43:31.85& $-$59:36:06.7& $-$31.9& none& \\
8&10:43:29.69& $-$59:35:53.7& $-$31.6& none& Bipolar, PA $\sim$ 20 degrees\\
9&10:43:32.00& $-$59:36:24.5& $-$5.4& proplyd& Teardrop-shaped extension at PA $\sim$ 305 degrees\\
\noalign{\smallskip}
\enddata
\end{deluxetable*}

The $^{12}$CO datacubes contain a dozen or so compact sources whose emission is
restricted to an interval of $\sim$ 1 km$\,$s$^{-1}$, and are clearly
distinct from the velocity fields associated with the spatially extended
molecular clouds (see Table~\ref{tab:compact}).
These sources are labeled with numbers in Fig.~\ref{fig:pos-vel}
with the exception of object 1, which is too blueshifted to be included in the figure.
Most of these sources appear to be individual globules situated
either foreground or background to the Western Wall cloud, but some may represent outflows.
Objects 3, 4, and 6 appear as small, bright-rimmed dark globules about 0.5 arcseconds (1150~AU) in
diameter superposed upon the Western Wall cloud in H$_2$ and Br-$\gamma$ adaptive optics images of the region
\citep{hartigan20}, while object 1 lies near a small arc of H$_2$
emission that may or may not be related to the CO source. 
Objects 6 and 8 have bipolar spatial morphologies in the $^{12}$CO cubes,
and exhibit velocity gradients suggestive of an outflow.
Object 2 has several knots arrayed along an arc with a
clear velocity gradient, while object 
9 is a teardrop-shaped cloud that resembles a proplyd, and 
glows faintly in H$_2$ but is invisible at Br-$\gamma$.
This latter object, situated to the east of the Western Wall cloud,
is the only source in Table~\ref{tab:compact} with a more redshifted radial
velocity than the Western Wall cloud. The tail of the proplyd is more redshifted than
the head, and breaks into three knots in its most redshifted velocity slice.
Object 9 is probably located behind the Western Wall cloud, as this source has similar radial
velocities to Cloud~B (Fig~\ref{fig:pos-vel}), which must
be a background object as noted above.
Some of the knot-like morphology we observe in Fig~\ref{fig:pos-vel}
on scales of the beam size (e.g. the string of knots that comprise
object 2) may be enhanced by noise or from uncertainties in the
reconstruction of the images from the raw ALMA data.
We defer additional discussion of these objects to future works.

\subsection{The Western Wall Molecular Cloud}
\label{sec:ww}

\subsubsection{Optical Depths, Depletion from Irradiation, and Cloud Mass}
\label{sec:opt-depth}

Maps of the $^{12}$CO, $^{13}$CO, C$^{18}$O, and [C~I] line peak intensity recorded in the
Western Wall velocity range are shown in Fig.~\ref{fig:co_isotope}.  The intensity in the maps
above the microwave background
is in units of the brightness temperature $T_b$, calculated using the Planck equation as

\begin{equation}
\label{eq:tb_p}
    T_b = \frac{h \nu}{k}\left[ \ln \left( \frac{2 h \nu^3 \Omega}{c^2 F_\nu}+1 \right) \right]^{-1}.
\end{equation}

\noindent
where $F_\nu$ is the specific flux integrated over a resolution element defined by
the solid angle $\Omega$ = $\pi \theta_{min} \theta_{max}$ / $(4 \ln 2)$, where
$\theta_{min}$ and $\theta_{max}$ are the minimum and maximum FWHM of the synthesized
beam.  The $^{12}$CO emission reaches a peak brightness temperature between 80~K and 90~K along the
photoevaporation front where the stellar radiation heats the
molecular cloud. 

The observed brightness temperatures are somewhat lower than those found in 
the Orion Bar ($\sim$~150~K) and NGC 7023 ($\sim$ 110~K), where the PDRs are
located closer to their respective photoionization sources
\citep{Joblin18}. For example, the distance between $\theta^1$ Ori C, the hottest
star in the Trapezium, and the main ionization front is about 0.25 pc, whereas
the Western Wall is about 2.3 pc in projected distance away from Tr 14. 
\cite{brooks03} lists the irradiation flux at the
Orion Bar as $\sim$ $4\times 10^4$ $G_0$ (where $G_0=1$ corresponds to a flux
of $1.6\times10^{-3}$ erg cm$^{-2}$ s$^{-1}$), while recent models of the Western Wall give a 
comparable value of $3\times 10^4$ $G_0$ \citep{Wu18}.
However, it is important to keep in mind that the Western Wall has $\sim$ 50 - 100 times the area
of the Orion Bar, so radiation can influence molecular clouds in Carina
on much larger scales than it can in less-massive regions such as Orion.

We can get some idea as to the optical depths we might expect for a
line by calculating the 
optical depth at line center for thermal broadening under typical molecular
cloud conditions and assuming standard abundances (see Eqn.~\ref{eq:taufinal} in the Appendix):

\begin{equation}
\label{eq:tau0}
\tau_0 = 1741\ \lambda^3A_{21}\frac{g_2}{g_1}m^{0.5}T^{-0.5}S
\frac{{\rm N}_1}{{\rm N}_{{\rm TOT}}}\frac{{\rm N}_{{\rm TOT}}}{{\rm N}_H}\frac{{\rm N}_{\rm H}}{10^{21}{\rm cm}^{-2}}
\end{equation}

\smallskip

\noindent
where $\lambda$ is the wavelength in microns, A$_{21}$ is the Einstein-A value
for the transition in s$^{-1}$, g$_2$/g$_1$ is the ratio of the statistical
weights between the upper and lower states, m is the species mass
in amu, T is the temperature in K, S is a factor that corrects for stimulated emission
and for departures from LTE, and the final three terms are, respectively,
the fraction of species in the lower level state, the abundance of the species respective
to hydrogen, and the hydrogen column density.
 
\begin{figure}[!ht]
\centering
\includegraphics[width=1.0\linewidth]{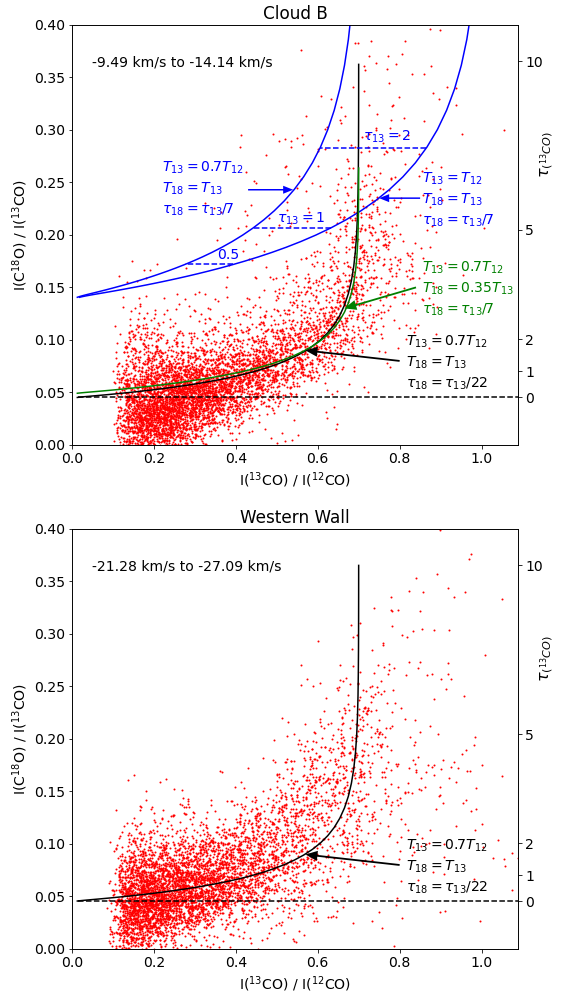}
\caption{Observed intensity ratios between
$^{12}$CO, $^{13}$CO, and C$^{18}$O for Cloud B (top)
and the Western Wall cloud (bottom). Ratios were taken from positions
separated by $\sim$ 1 beam size within each 0.166 km$\,$s$^{-1}$
velocity channel spanned by the cloud, and are shown for all points in the
datacube where the $^{13}$CO intensity exceeds 0.25 Jky/beam.
Optical depths in $^{13}$CO for the black curve are shown on
the scale at right. The simple models shown are discussed in the text.
}
\label{fig:cotau}
\end{figure}

Taking T $\sim$ 30~K, an abundance ratio 
C~I/CO $\sim$ 0.2 in the PDR with 40\%\ of C locked in grains
(see Appendix~C, Table~\ref{tab:tau}, and discussion below), and a fiducial hydrogen 
column density of 2.7$\times$10$^{21}$ cm$^{-2}$, which corresponds to a visual extinction
A$_V$ $\sim$ 1 for a normal reddening law \citep{Liszt14}, we find
C$^{18}$O is optically thin with $\tau$ $\sim$ 0.2 (with no radiative depletion), and
[C~I] should have an optical depth $\sim$ 0.8. 
$^{13}$CO should be more optically thick with $\tau$ $\sim$ 1.3,
and $^{12}$CO has a very high optical depth of $\tau$ $\gtrsim$ 100.
These numbers are meant only as guides to help interpret the images,
as following the layered distribution of molecules, atoms, ions, and dust within
a PDR is a complex problem that requires sophisticated
modeling \citep[e.g.][]{Spaans96}. Nonetheless, overall we expect $^{12}$CO to be very 
optically thick and to probe only the outer layer of the cloud, $^{13}$CO to be
either optically thick or thin depending on the region, 
[C~I] to be mostly optically thin except perhaps in the dense cores, and
C$^{18}$O to be optically thin everywhere. These expectations are
in agreement with the results in Fig.~\ref{fig:cotau} described in Sec.~\ref{sec:opt-depth}.
Optically thin lines tracers such as C$^{18}$O J = 2 - 1, and, to
a lesser extent [C~I] 609~$\mu$m, probe
the densest portions of the clouds.  Of the two tracers, C$^{18}$O has
the better signal-to-noise, as its J = 2 - 1 1.3~mm line emits in ALMA's Band 6,
in contrast with the more difficult observation of [C~I]~609$\mu$m
in ALMA's Band 8.

The line optical depth $\tau_\nu$ for a slab of gas in LTE at temperature T
depends upon the observed specific line intensity I$_\nu$ via

\begin{equation}
\label{eq:I-tau}
{\rm I}_\nu = {\rm B}_\nu({\rm T})(1-e^{-\tau_\nu}), 
\end{equation}

\noindent
where B$_\nu$ is the Planck function. With assumptions of constant temperature
and fixed abundance ratios between 
$^{12}$CO, $^{13}$CO and C$^{18}$O, the observed line ratios between these
isotopologues in principle provide an optical depth map for the region at each velocity.

Unfortunately, this simple procedure fails for two reasons. First, the inner regions
of molecular clouds are more shielded from external radiation, and thus are colder than
the outer layers of the cloud. Because $\tau$ $\sim$ 1 occurs closer to the surface
of a cloud for $^{12}$CO than it does for a less abundant isotopologue such as $^{13}$CO, the
intensity of $^{12}$CO can be significantly higher than that of $^{13}$CO even when
both lines are optically thick. Fig.~\ref{fig:co_isotope} illustrates this effect,
where the peak brightness temperature in $^{12}$CO is about 80~K, while that in 
$^{13}$CO is $\sim$ 50~K.

An even larger error is introduced by assuming constant abundance ratios for
$^{12}$CO : $^{13}$CO : C$^{18}$O throughout the cloud. As described by \citet{miotello14},
the abundance ratio of, for example, C$^{18}$O / $^{12}$CO can be depleted by
up to a factor of 20 within irradiated clouds owing to
processes that lead to selective dissociation of the different CO isotopologues. 
Photodissociation of CO is dominated by absorption into discrete bands above
the CO dissociation energy of $\sim$ 11.1~eV and below the H-ionization limit of 13.6~eV
\citep[e.g.][]{coband}, and
the energy levels of these bands differ enough between C$^{18}$O and $^{12}$CO,
that $^{12}$CO effectively self-shields against dissociation close to the surface
of the cloud without also shielding C$^{18}$O. As a result,
C$^{18}$O is dissociated much deeper into the cloud \citep{bally82,visser09}.

%
%
%

Fig.~\ref{fig:cotau} displays the observed line ratios between the three isotopologues
of CO for the velocity range and spatial extent of Cloud B, located along
the northern boundary of our map (Fig.~\ref{fig:pos-vel}) and for the Western Wall cloud.
Cloud B is defined by a 40\arcsec\ $\times$ 30\arcsec\ box centered at
$\alpha$(2000) = 10:43:31.9, $\delta$(2000) = $-$59:35:40.4
and encompassing $-$9.49 km$\,$s$^{-1}$ to 
$-$14.14 km$\,$s$^{-1}$, and the Western Wall cloud by
a 36\arcsec\ $\times$ 72\arcsec\ box centered at
$\alpha$(2000) = 10:43:30.0, $\delta$(2000) = $-$59:35:55
between $-$21.28 km$\,$s$^{-1}$ and $-$27.09 km$\,$s$^{-1}$.
Cloud B is in the Carina Nebula, as it 
appears weakly in H$_2$ (Fig~\ref{fig:composite}).
However, its relative faintness in H$_2$ and lower peak
temperature in $^{12}$CO ($\sim$ 60~K instead of $\sim$ 80~K for
the Western Wall cloud) implies that the radiation field impinging
upon Cloud~B is substantially lower than it is for the Western Wall.
Hence, any differences in the emission line ratios between
these two clouds informs how radiation
affects the isotopologues of CO in molecular clouds.

The blue curves in the diagrams illustrate the expected line ratios for the simplest
case of fixed cosmic abundances and temperatures, where N($^{13}$CO) / N(C$^{18}$O) = 7
and N($^{12}$CO) / N($^{13}$CO) = 77 \citep{Wilson94}. The model with equal
temperatures for $^{12}$CO and $^{13}$CO does not agree with the observations.
Adopting T$_b$($^{13}$CO) = 0.7T$_b$($^{12}$CO) to account for the warmer temperatures 
associated with the $^{12}$CO emission near the surface of the cloud reproduces 
the asymptotic behavior of the I($^{12}$CO) / I($^{13}$CO) ratio 
where the regions are brightest and most optically thick, but this model
fails to account for the anomalously low I(C$^{18}$O) / I($^{13}$CO) ratio
present in both clouds. 

The black curve is a model calculated by reducing the  
abundance of N(C$^{18}$O) / N($^{13}$CO) from 1/7 to 1/22, and this model
matches the data reasonably well for Cloud B albeit with significant scatter.
Adopting a greatly reduced temperature in the C$^{18}$O emitting region
relative to the temperature in the $^{13}$CO region (green curve) agrees equally
well with the data, however this model would imply T(C$^{18}$O) $\sim$ 0.25 T($^{12}$CO)
$\lesssim$ 20~K, which seems rather low.
The scatter at the left side of the plot arises primarily
from low flux values, especially in the C$^{18}$O map, while the scatter in the
upper right probably arises mostly from variations in the temperature ratio between
$^{12}$CO and $^{13}$CO across the region.

Although the irradiated Western Wall cloud generally follows the same trends as we see in Cloud B,
the overall fit of the model to the data is not as good, most noticably where
the data congregate above the black curve as the curve
begins to bend upward. The differences
between the model and Western Wall data are in the sense that
C$^{18}$O appears more abundant or warmer in the data relative to the 
model when $\tau$($^{13}$CO) $\sim$ 2. Regardless of the cause, there
is a systematic difference between the isotopologue CO ratios in the
highly-irradiated Western Wall cloud, and the more weakly-irradiated Cloud B.

\begin{figure}[!t]
\centering
\includegraphics[width=1.0\linewidth]{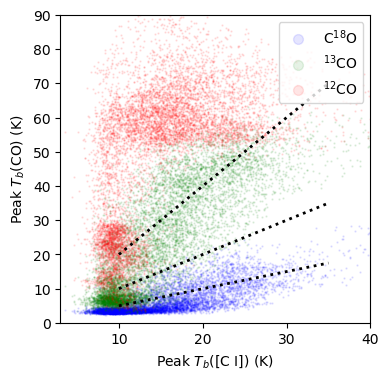}
\caption{Peak brightness temperatures of the J = 2 - 1
transitions of $^{12}$CO (red), $^{13}$CO (green), 
and C$^{18}$O (blue) plotted against the peak brightness
temperature of [C~I] 609 $\mu$m within the Western Wall
molecular cloud. The dashed lines are fiducials with
slopes of 0.5, 1, and 2. There is a nearly linear
relationship between the peak brightness temperatures
of the [C~I] and C$^{18}$O lines.}
\label{fig:C1vsCO}
\end{figure}

The axes on the right side of the plots in
Fig.~\ref{fig:cotau} show that $^{13}$CO ranges in optical depth
within each velocity channel
from optically thin to a maximum of $\sim$ 10. Hence, 
C$^{18}$O is optically thin throughout the region, with an
optical depth $\tau_{18}$ $\lesssim$ 0.5, and
we can integrate the observed line flux to estimate
a mass for the Western Wall cloud.
The total flux of the line integrated over the Western Wall cloud between 
$-$27.09~km$\,$s$^{-1}$ and $-$21.28~km$\,$s$^{-1}$
is $1.7\times 10^{-15}$ erg$\,$cm$^{-2}$s$^{-1}$    
with an error of 10\%\ dominated by the systematic errors associated
with the absolute flux calibrations in the total power data.
At a distance of 2.3 kpc this flux translates to a line luminosity of
L$_{21}$ = $1.1\times 10^{30}$erg$\,$s$^{-1}$. The line luminosity relates to
the total number of C$^{18}$O molecules in the J = 2 level N$_2$(C$^{18}$O) via

\begin{equation}
{\rm N}_2({\rm C}^{18}{\rm O}) = \frac{{\rm L}_{21}}{{\rm A}_{21}{\rm h}\nu} = 1.26\times 10^{51},
\end{equation}

\noindent
where we have used A$_{21}$ = $6.01\times 10^{-7}$ s$^{-1}$ as the Einstein-A coefficient
and h$\nu$ = $1.45\times 10^{-15}$ erg as the transition energy. 

For a characteristic temperature of 30~K and LTE population, 25\%\ of the
C$^{18}$O molecules are in level 2, so adopting a conversion factor of 22$\times$77 = 1694 between
the abundances of C$^{18}$O and $^{12}$CO implies $8.5\times 10^{54}$ $^{12}$CO molecules.
The abundance ratio between of H$_2$ and gaseous CO in
star-forming regions varies between $\sim$ 3100 and 14500 \citep{lacy94,lacy17}. Adopting 6000
for this value we calculate the mass of the portion of the Western Wall cloud in our ALMA datacube to be
85~M$_\odot$. The same calculation done with $^{13}$CO assuming a ratio of $^{12}$CO/$^{13}$CO
= 77 gives 52~M$_\odot$ for the mass, a lower value as expected given that much of the 
$^{13}$CO is optically thick.

Based on their lower-resolution $^{12}$CO and $^{13}$CO maps in the velocity range
between $-$25~km$\,$s$^{-1}$ and $-$22~km$\,$s$^{-1}$,
\citet{brooks03} found a much higher mass of $\sim$ 500 M$_\odot$ assuming either
virial equilibrium or using a conversion from $^{13}$CO to H$_2$ applicable if the $^{13}$CO
is optically thin. However, these estimates cover an area five times that of our map,
and virial calculations will overestimate the masses in this region owing to the
multiple velocity components along the line of sight.
Fig.~\ref{fig:composite} also suggests that the
entire Western Wall cloud contains several dense regions that are not included in our
ALMA data. Overall, the Western Wall cloud is larger in area by a factor of $\sim$ 2 - 3
than our mapped region, so our best estimate for
the total mass of the entire cloud is $\sim$ 240~M$_\odot$, though this number could change
by a factor of two with more extensive mapping of the region. The uncertainties in
these mass estimates are dominated by systematic errors associated with the H$_2$/CO
conversion factors and the abundance ratios of the isotopes of CO, and are likely
to be a factor of two given the observed range of these values between different
molecular clouds.
 
\subsubsection{{\rm [C~I]} in the Western Wall Cloud}
\label{sec:ci}

In the Orion A and Orion B PDRs, there is a correlation between the brightness temperature
of the [C~I] $^3$P$_1-^3$P$_0$ and that of $^{13}$CO J = 1 - 0 line, with $T_b(\textrm{[C~I]})
\sim 0.6 \times T_b(\textrm{$^{13}$CO})$ \citep{Ikeda99, Ikeda02}.
The intensity ratio indicates that [C~I] 609$\mu$m has
an optical depth that varies between about 0.3 and 2 and implies
C~I column densities between $10^{17}$ and 10$^{18}$ cm$^{-2}$. These studies found that
the abundance ratio between atomic carbon and CO molecules is almost constant
between 0.1 and 0.2 across regions with different levels of star formation activities,
such as Orion KL and the much more quiescent L1641 dark cloud. Hence, the C~I/CO ratio is
insensitive to the level of UV radiation, at least in Orion. 

The ratio between the intensities of the CO and the C~I
lines measured toward the Western Wall region is shown in Fig.~\ref{fig:C1vsCO}.
Unlike \cite{Ikeda02}, our observations targeted
the J = 2 - 1 transition of CO, covered three CO isotopologues,
and resolved emission on spatial scales almost 30 times smaller
than those observed in Orion. Despite the observational differences,
we find a similar correlation between [C~I] and CO line
intensities. As expected for optically thin lines with an approximately
constant abundance ratio, the correlation between [C~I] and
C$^{18}$O is nearly linear, with the brightness temperature of
[C~I] 609 $\mu$m about twice that of C$^{18}$O J = 2 - 1.
However, there is scatter in the relationship and there
are features that are unique to the [C~I]
map. For example, in Fig.~\ref{fig:co_isotope} the
[C~I] emission lacks the bright, diffuse features present in the center
of the C$^{18}$O map. [C~I] also shows additional
small droplet-shaped overdensities with clear velocity gradients
that are less apparent in C$^{18}$O, and are 
approximately the same size of the beam ($\sim
1\arcsec$).

\begin{figure}[!t]
\centering
\includegraphics[width=1.0\linewidth]{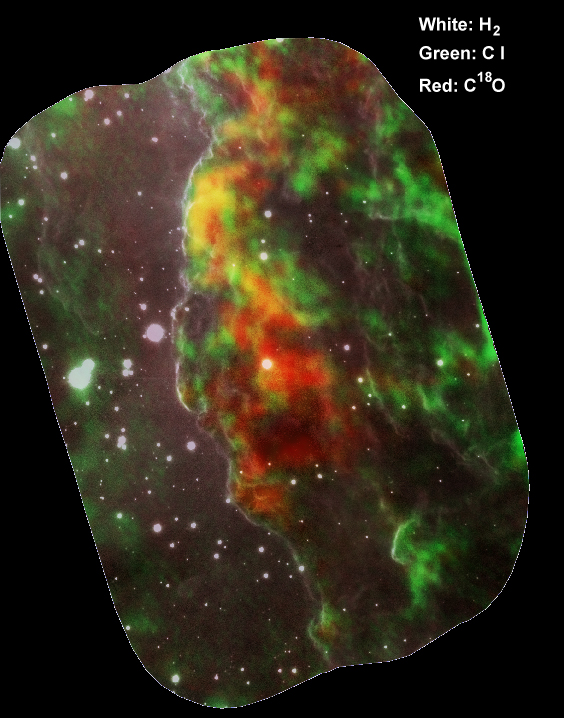}
\caption{Color composite of H$_2$, [C~I], and C$^{18}$O in the
Western Wall molecular cloud. The H$_2$ image in white is from
Gemini South's adaptive optics imager \citep{hartigan20}, and
shows fluorescence at the boundary of the cloud.
The [C~I] (green), and C$^{18}$O (red) images extracted from the ALMA
datacubes between V$_{lsr}$ velocities $-$21.28 km$\,$s$^{-1}$
and $-$27.09 km$\,$s$^{-1}$ are mostly optically thin. There is a
general progression from H$_2$ to [C~I] to C$^{18}$O as 
one moves deeper into the cloud from left to right.
}
\label{fig:h2c1c18}
\end{figure}

\begin{figure*}[!ht]
\centering
\includegraphics[width=0.9\linewidth]{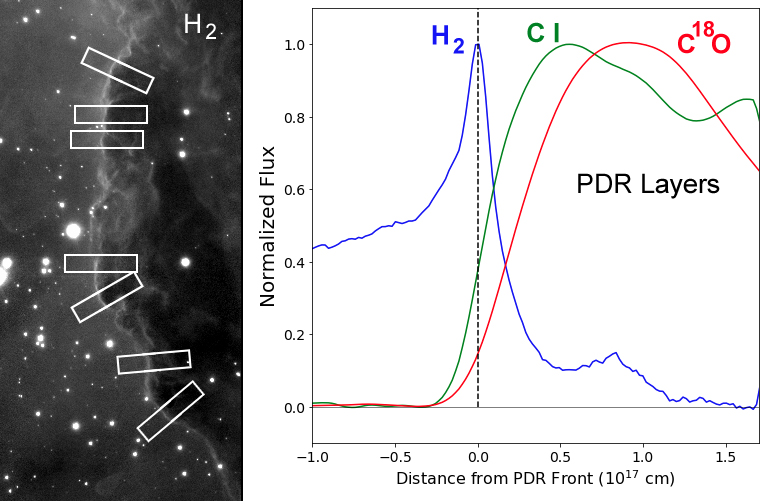}
\caption{
Average spatial offsets between H$_2$, C~I, and C$^{18}$O emission in the Western
Wall molecular cloud. The average spatial line profiles shown in the graph at right
were extracted from the seven boxed areas marked 
in the H$_2$ image from \citet{hartigan20} at left. The H$_2$ image includes 
continuum, which produces the plateau of emission upstream (to the left) of 
the PDR. The C~I and C$^{18}$O profiles were restricted to the velocity range 
associated with the cloud.}
\label{fig:offsets}
\end{figure*}

While there is a positive correlation between
[C~I] 609 $\mu$m and both $^{13}$CO J = 2 - 1 and $^{12}$CO J = 2 - 1 in
Fig.~\ref{fig:C1vsCO}, the relations do not fall on a line, especially
for $^{12}$CO, which is very optically thick so its
brightness temperature saturates at the thermal temperature of the
gas.  As in Orion, the peak brightness temperature
of [C~I] 609 $\mu$m is about half of that of the $^{13}$CO J = 2 - 1
over most of the Western Wall cloud.

\subsubsection{Observable PDR Layers}
\label{sec:pdr}

Our ALMA data are ideal for illustrating the overall layered structure
of a PDR because we observe both [C~I] and C$^{18}$O, and the C$^{18}$O is
optically thin and the [C~I] mostly so. 
[C~I] is predicted to have an emitting layer that
lies closer to the PDR front than CO does \citep{Spaans96}. Moreover, the
Western Wall has a convex shape and we view it in profile, which greatly reduces
problematic projection effects that complicate images of concave cavities
such as the Orion Bar. Fig.~\ref{fig:h2c1c18} overlays the latest high-resolution
adaptive-optics image of the region in H$_2$ \citep{hartigan20} with the [C~I] and C$^{18}$O maps.
As predicted by theory, H$_2$ traces the interface where radiation 
impinges upon the cloud. The interface is followed by a complex flocculent morphology
in the ALMA images, but there is order in the sense that all the bright
C$^{18}$O features have a layer of [C~I] between the CO and the H$_2$.

\begin{figure*}[!t]
\centering
\includegraphics[width=0.8\linewidth]{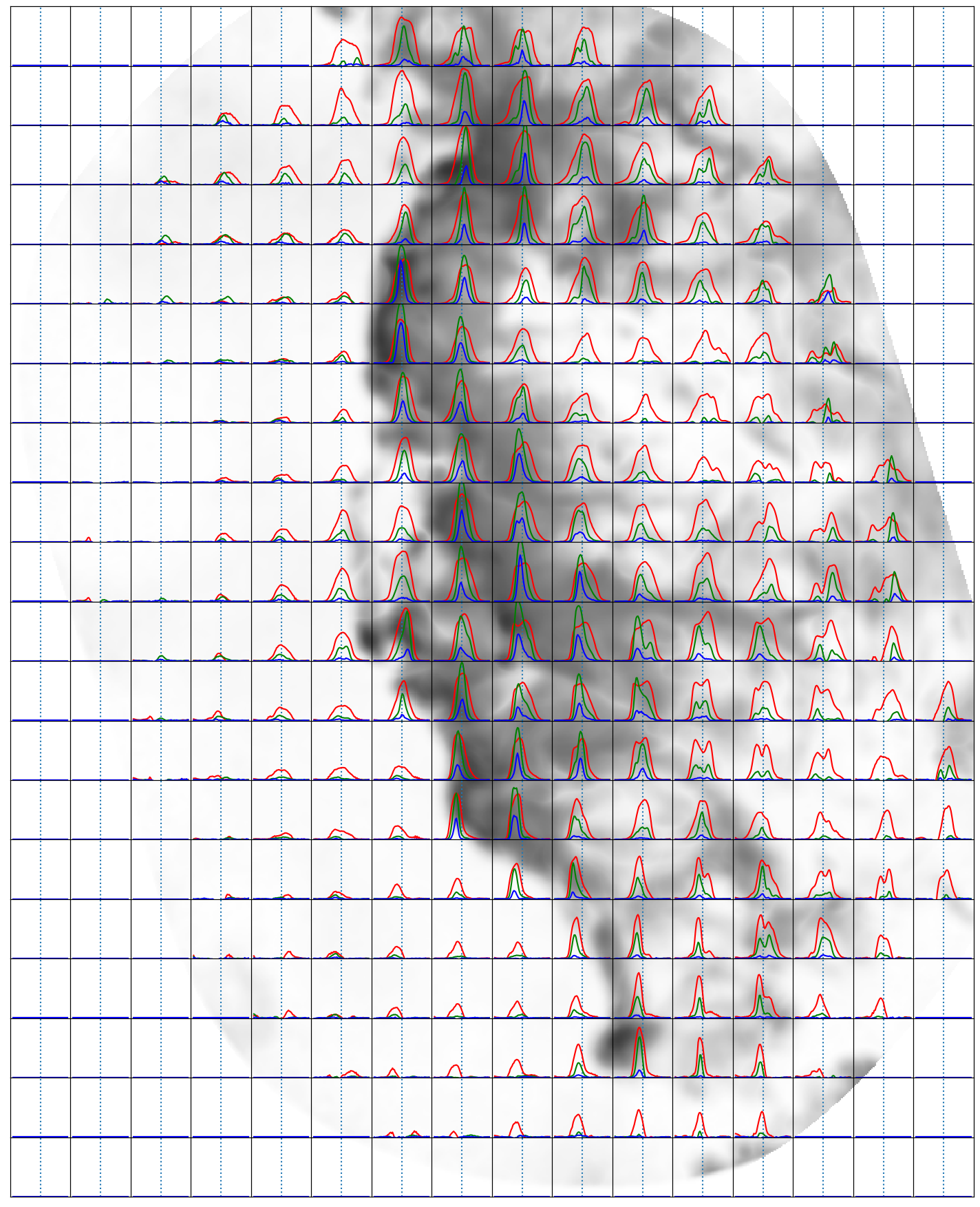}
\caption{
Spectra of the $^{12}$CO (red), $^{13}$CO (green), and
C$^{18}$O (blue) J = 2 - 1 lines recorded toward the Carina Western Wall molecular cloud.
For a better comparison, the intensity of the $^{13}$CO and C$^{18}$O lines were multiplied by a
factor of 2 and 5, respectively. The spectra were obtained by spatially integrated
the observed line emission across square regions of 4\arcsec$\times$4\arcsec,
and span velocities between -29.59 km s$^{-1}$ to $-19.62$ km s$^{-1}$, from left to right in each
box. As a reference, vertical dotted lines indicate the average velocity, -24.60 km s$^{-1}$ of the cloud.
The background image shows the peak intensity map of $^{13}$CO.  The wider and
multi-peaked spectra in the cloud's interior are likely due to multiple clumps
overlapping along the line of sight. Conversely, the clumps located along
the Western Wall's edge have spectra with a single peak.}
\label{fig:co_spec}
\end{figure*}

The offsets are perhaps easiest to see in Fig.~\ref{fig:offsets},
which combines the spatial emission profiles along seven 
transects, each 2\arcsec\ in width, chosen to contain bright emission and
to avoid stars.  The excellent spatial resolution of the H$_2$ adaptive-optics
image helps greatly to reduce stellar contamination in these profiles,
but continuum light adds to the integrated flux in the H$_2$ profiles,
especially east of the PDR in the direction of the radiation sources.
Nonetheless, the integrated profiles display a sharp H$_2$ peak, which
we take to define the location of the PDR. The profiles in 
Fig.~\ref{fig:offsets} exhibit the expected layered structure,
with offsets between the H$_2$,
C~I, and C$^{18}$O emitting layers of $\sim$ $10^{17}$ cm, similar to
the offsets observed between H$_2$ and Br-$\gamma$ in this region by
\citet{carlsten18}.

\subsubsection{Kinematics}
\label{sec:kinematics}

Figure~\ref{fig:co_spec} depicts the complex kinematics within the Western Wall molecular cloud in the lines
of $^{12}$CO, $^{13}$CO and C$^{18}$O. Regions close to eastern edge where the
Western Wall is irradiated have relatively narrow, single-peaked lines, whereas
regions away from the photo-evaporation front exhibit wider, and sometimes multiple-peaked profiles.
These multiple peaks make sense from a geometrical standpoint, as our line
of sight passes through more of the cloud to the west as the projected distances from
the front increase, so more clumps will appear within the beam if the clumps
are distributed through the cloud.
The line kinematics uncovered a large-scale velocity gradient south-to-north across
the Western Wall cloud, with line emission from the north part of the object peaking at a velocity of
about $-$24 km$\,$s$^{-1}$ and the south part peaking at velocities of $-$26 km$\,$s$^{-1}$.
This gradient is particularly striking in the animation (Fig.~\ref{fig:movie}).
We defer further analysis of the kinematics within the Western Wall cloud and its connection
to turbulence to a companion paper \citep{downes22}.

\subsection{Comparing CO and Continuum Maps, Distortion of Cores at the PDR, Dust Properties}
\label{sec:cont}

\begin{figure*}[!t]
\centering
\includegraphics[width=1.0\linewidth]{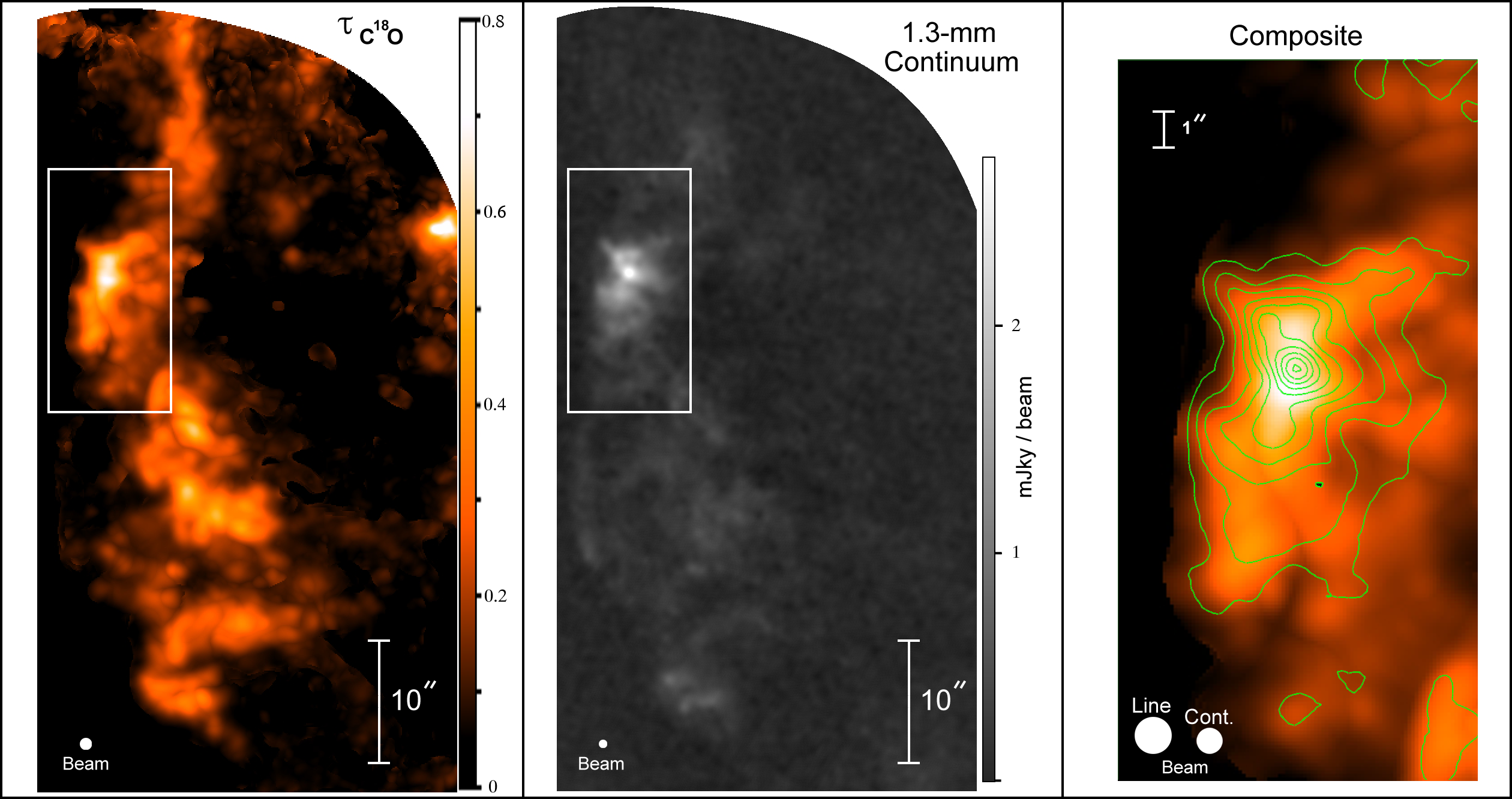}
\caption{
Left: Map of the peak optical depth in C$^{18}$O for the Western Wall Cloud.
The boxed region centered at
$\alpha$ (2000) = 10:43:31.6 $\delta$ (2000) = $-$59:35:40.0
subtends 10\arcsec\ $\times$ 20\arcsec , and contains several 
bright resolved structures.
Center: 1.3~mm continuum map, acquired with a circular 0.7\arcsec\ beam. 
There is a bright unresolved core near the location of peak
C$^{18}$O optical depth.
Right: Expanded view of the boxed
area. The color representation is the same C$^{18}$O 
optical depth map at left, and contours in units of 0.3 mJy/beam
depict the 1.3~mm continuum. The C$^{18}$O is somewhat
flattened along the PDR interface.
}
\label{fig:cotaucont} 
\end{figure*}

Eqn.~\ref{eq:I-tau} provides an optical depth
measurement at each position and velocity from the observed intensity ratio
between $^{13}$CO and C$^{18}$O, under the assumption that the two species
emit at the same temperature.
Because all areas with substantial C$^{18}$O flux are
optically thick in $^{13}$CO (Fig.~\ref{fig:cotau}), this measurement
is independent of the relative abundances of $^{13}$CO and C$^{18}$O.
The left panel of Fig.~\ref{fig:cotaucont} shows the maximum optical depth
achieved across the datacube for the Western Wall cloud.  Interestingly, regions with
higher optical depth are aligned along the Western Wall, suggesting 
that the photoevaporation front is responsible for compressing the
molecular gas along this interface. In particular, the brightest knots
within the boxed area tend to curve in the same direction as the
PDR interface does.
However, the flattened structures in C$^{18}$O do not resolve into
separate pillars, as should occur when a radiation front wraps
around dense knots.

The flattening of dense structures along the PDR is
somewhat less evident in the 1.3-mm continuum observations in the middle
panel of Fig.~\ref{fig:cotaucont}. The continuum has an
unresolved point source at 10:43:31.60 $-$59:35:38.6 embedded within
an extended structure and surrounded by dense C$^{18}$O knots.
The brightness temperature
of the continuum emission in Fig~\ref{fig:cotaucont} reaches maximum values above
the microwave background of about
2.5~K at 1.33~mm and 5.6~K at 0.62~mm. These temperatures are lower than those registered
in the C$^{18}$O emission line, and indicate that the continuum is optically
thin. If the gas and dust temperatures are 30~K, the optical depth of the
1.33~mm continuum is $\sim$ 0.1, and the optical
depth of the 0.62~mm continuum is $\sim$ 0.2. 

Because the continuum at 1.3-mm is optically thin, the observed total
specific flux integrated over the Western Wall cloud of 280 mJy 
provides a mass estimate for the cloud.  Using the relation

\begin{equation}
\label{eq:dustmass}
{\rm F}_\nu = {\rm B}_\nu({\rm T})M\kappa/{\rm d}^2
\end{equation}

\noindent
and taking $\kappa$ = 0.006 cm$^2$g$^{-1}$ as the opacity
\citep[this value includes the gas/dust ratio of 100][]{Beckwith91} 
the inferred mass is 31~M$_\odot$ for T = 30~K and 50~M$_\odot$ for T = 20~K.
These values agree to within a factor of $\sim$ 2 with those measured from the C$^{18}$O observations in 
Sec.~\ref{sec:opt-depth}.
The total 1.3-mm continuum in the boxed region in Fig.~\ref{fig:cotaucont}
is 148~mJy at 225~GHz, which translates to 27 M$_\odot$ if T = 20~K 
or 16~M$_\odot$ if T = 30~K using Eqn.~\ref{eq:dustmass}.

\begin{figure*}[!t]
\centering
\begin{interactive}{animation}{gaussclump.mp4}
\includegraphics[width=1.0\linewidth]{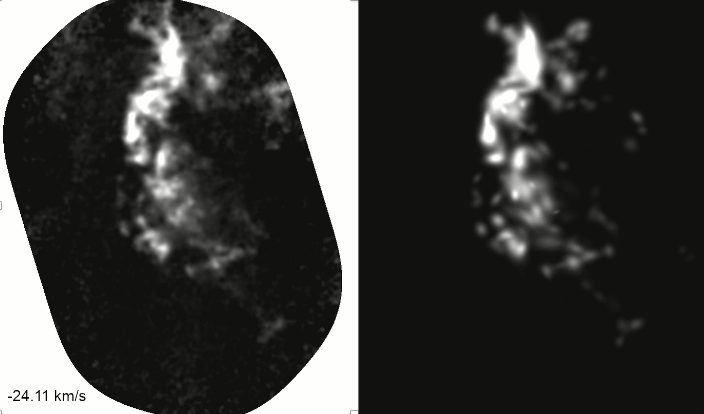}
\end{interactive}
\caption{Animation through the velocity frames of the C$^{18}$O ALMA datacube
of the Western Wall Cloud (left) and the corresponding model produced by fitting
254 Gaussian clumps to the data.
The animation moves through each frame between $-$26.76 km$\,$s$^{-1}$ and $-$22.61 km$\,$s$^{-1}$
in 0.17 km$\,$s$^{-1}$ intervals.  The video duration is 23 seconds.
}
\label{fig:gaussmovie}
\end{figure*}

The signal-to-noise in the 0.62 mm continuum map is only high enough to estimate
a spectral slope of the continuum between 1.33 mm and 0.62 mm
over a limited area in our map, but this measurement is useful because it
constrains the sizes of the dust grains responsible for the emission. 
The intensity of the dust continuum emission
follows $I_\nu = B_\nu(T)(1-e^{-\tau_{\nu}})$, where the dust
optical depth $\tau_\nu$ is the product of the dust column density $\Sigma$ and
the dust opacity $\kappa_\nu$ = $\kappa_0 (\nu/\nu_0)^\beta$, where
$\beta$ is a parameter that depends upon the dust grain size and
composition \citep{Draine06}. If the observed spectral index of
the intensity is $\alpha$ so that I$_\nu$ $\sim$ $\nu^\alpha$,
then for $\tau \ll 1$ as it is here we can estimate $\beta$ as

\begin{equation}
\label{eq:beta}
    \beta = \alpha - \frac{\ln{(B_{\nu 1}(T)/B_{\nu 2}(T))}}{\ln{(\nu 1/ \nu 2)}}.
\end{equation}

For $\nu_1$ = 228~GHz and $\nu_2$ = 479~GHz, 
the spectral index $\alpha$ in our maps ranges between about 3.4 $-$ 3.75. 
If the dust grain temperature is between 20~K and 50~K,
Eqn.~\ref{eq:beta} implies 1.5 $\lesssim$ $\beta$ $\lesssim$ 2.2. This value
is similar to the index of 1.7 for interstellar grains, implying that the grains are
smaller than 10 $-$ 100 $\mu$m \citep[see, e.g, Fig.~4 in][]{Testi14}. 
Hence, our data are consistent with the
properties of interstellar dust, and provide no evidence 
for grain growth in the Carina Western Wall cloud on the $\sim$ 2300~au
spatial scales probed by the observations.

\section{C$^{18}$O Clumps}
\label{sec:c18oclumps}

In this section we focus on identifying and characterizing the densest
regions in our maps as revealed by the most optically thin line tracer at hand,
the C$^{18}$O J~=~2~-~1 emission line. In observational papers such as ours,
it is standard practice to use the term `core' to identify a localized peak
in a mm-continuum map \citep[e.g. Section 4.4 of][]{AquilaCores}. Because the 
signal-to-noise is lower in our continuum maps than it is in our datacubes,
we focus primarily on the datacubes to identify compact structures.  We call the
localized emission peaks in our C$^{18}$O cube `clumps'. `Clump' is a flexible
term star-formation researchers have used to refer to structures as large as a parsec
\citep{PerseusCores}, or as small as a few hundred au \citep{kramer98}, depending on the
instrumental resolution.  In our maps, the C$^{18}$O clumps are typically
of order 0.01~pc. \citet{droplets} identified small ($\sim$ 0.05~pc) 
coherent gas structures similar to our C$^{18}$O clumps and called them `droplets',
with the definition that droplets must be pressure-confined and unbound gravitationally.
As it is unclear whether a potential droplet is bound or unbound {\it a-priori},
we use the term `clump' for these features, and assess their dynamical states later.

Fig.~\ref{fig:cotaucont} shows there is a strong spatial correlation between 
the continuum cores and the C$^{18}$O datacube clumps,
though the peaks in each map do not always coincide.
Once we identify a clump, we use the line luminosity in C$^{18}$O to estimate its mass,
and spectra of clumps measure both their internal velocity
dispersion and the spatial distribution of clump velocities in the cloud.
Properties of the dust in the continuum cores were described in Sec.~\ref{sec:cont}.

Using clumps rather than cores has some benefits in identifying dense structures
that may form stars in the future. 
Unlike continuum core masses, C$^{18}$O clump masses are
independent of the gas/dust ratio and the dust opacity law.
In addition, because the C$^{18}$O map is a data cube,
it separates any spatially superposed dense concentrations with differing radial velocities,
something not possible to do with continuum maps. Our ALMA C$^{18}$O
datacubes benefit from having higher signal-to-noise ratios than are present in the
continuum. A drawback of using C$^{18}$O to trace mass is that CO
should freeze out onto grains when temperatures fall below $\sim$ 20~K,
as typically occurs in the outer envelopes of protostars \citep{tafalla02, tycho21}.
However as we discuss in
Sec.~\ref{sec:protostars}, the strong external radiation fields
throughout the Western Wall region
should keep most of the cloud material in our ALMA maps
above the CO freeze-out temperature, so in our case the C$^{18}$O clumps
are a reasonable proxy for the mass.

\subsection{Identifying Clumps in the C$^{18}$O Datacube}
\label{sec:c18oclumpid}

Astronomers have used several methods to identify clumps within datacubes, including
the algorithm Clumpfind, which embeds peaks within progressively fainter intensity
contours to find clump boundaries \citep{clumpfind}, fitting 
Gaussian peaks within a cube \citep{stutzki90}, the Fellwalker 
and Reinhold algorithms of
peak identification \citep{cupid}, and more advanced statistical constructs such
as dendrograms which uncover the heirarchy of clustering spatial scales within a
data set \citep[e.g.][]{roso08,williams19,takemura21}. Based on numerical tests,
\citet{li20} concluded that the Fellwalker, Gaussclump, and Dendrogram
algorithms exhibited the best overall performance when tested by their
ability to extract locations of known clumps within synthetic, noised datacubes.

It is important to keep in mind the final goal of the analysis, which is to identify
mass concentrations that might later form a single star or perhaps a multiple star
system in order to compare with the initial mass function for stars.
Fractal-like structures with irregular shapes are not well-suited for such comparisons. 
In our datacube we found that Clumpfind often did not make the 
choices we would have when it came to breaking up irregularly-shaped bright clumps
and it sometimes produced convoluted clump shapes. Overall, the
C$^{18}$O data cube largely consists of elliptically-shaped features in most
of the velocity slices, so it makes sense to use a relatively simple shape
such as a three-dimensional Gaussian to fit the data. 

Table~\ref{tab:clumps} compiles
the properties of 254 Gaussian clumps identified using the fitting algorithm of
\citet{stutzki90} as implemented by the CUPID package \citep{starlink,cupid}.
The algorithm uses several parameters to distinguish clumps
from noise, manage clump identification near the boundary of the map, define a
minimum clump size, and terminate the search once the clump model matches the
data sufficiently well \citep[e.g.][]{kramer98}.  Different model runs of the Gaussian fitting procedure
produced different numbers of clumps, the main changes
coming at the low-mass end where the algorithm is trying to decide if faint
residual structures in the map deserve to be identified as clumps, and when
the algorithm attempts to fit multiple Gaussians in an effort to match a
non-Gaussian shape. The best way to decide if a model has missed any major features
in the data or if it has overfit the data with too many clumps is simply to
display the observed cube alongside the model cube and step through both
in velocity. We present our best results with this type of animation in Fig.~\ref{fig:gaussmovie},
so the reader can assess the overall performance of the model. In these models
the minimum clump mass is 0.01 M$_\odot$.

\subsection{Clump Masses, Sizes and Internal Dynamics}
\label{sec:core_prop}

To find the mass of an individual clump, we use the procedure outlined in
Sec.~\ref{sec:opt-depth} to convert C$^{18}$O line luminosities to masses. 
These calculations give the total mass of the Western Wall molecular cloud to be 85~M$_\odot$.
The 254 clumps detected in the C$^{18}$O datacube range in mass from
0.011 M$_\odot$ to 4.93 M$_\odot$, and together account for 75 M$_\odot$ in the mapped area.
The 50 most massive clumps make up 55.3 M$_\odot$, or 74\%\ of the total mass
in clumps, while clumps below the median of 0.091 M$_\odot$ make up only 6.5 M$_\odot$, or 9\%\
of the total.

\begin{figure}[!b]
\centering
\includegraphics[width=1.0\linewidth]{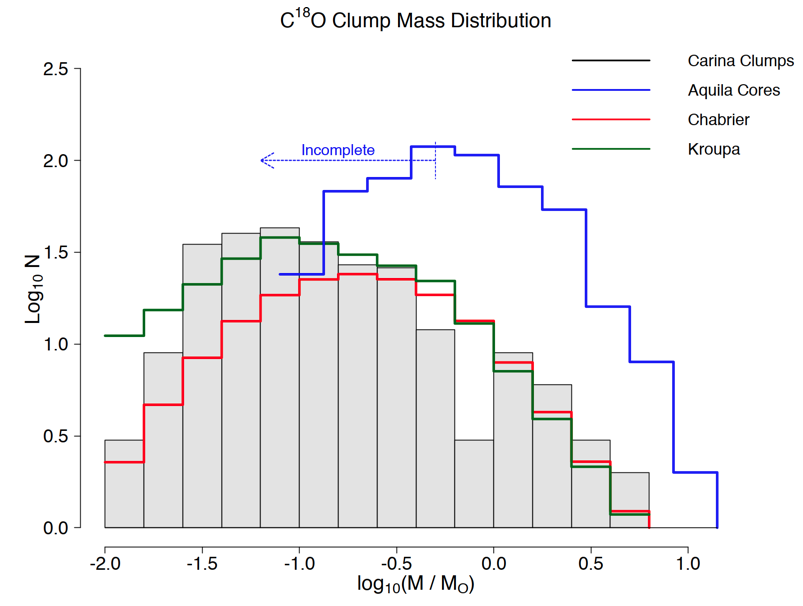}
\caption{
Observed clump mass distribution in the Western Wall molecular cloud plotted as a grey histogram.
The blue histogram is the core mass distribution found
by \citet{AquilaCores} in Herschel data of the Aquila Rift clouds (their Fig.~16).
The red and green histograms correspond to 75 M$_\odot$ of material distributed among stars with
M $<$ 6.31 M$_\odot$ according to the IMFs of \cite{chabrier05} and \citet{kroupa01},
respectively.
}
\label{fig:imf} 
\end{figure}

Fig.~\ref{fig:imf} shows that the clump masses in the Western Wall generally follow the
Chabrier \citep{chabrier05} and Kroupa \citep{kroupa01} IMFs,
with a bit of a deficit at masses just below 1 M$_\odot$.
The lowest clump masses match the Kroupa IMF somewhat better than the Chabrier IMF.
As noted above, the number of very low mass ($\lesssim$ 0.05 M$_\odot$) clumps
identified by Gaussclumps depends upon the parameters it uses, with more of these
found with a higher number of iterations as the code endeavors to fit Gaussians
to non-Gaussian shaped objects. This uncertainty should be kept in mind when evaluating
the numbers in the lowest histogram bins in Fig.~\ref{fig:imf}. 


Like the IMF, the clump mass distribution in the Western Wall is shifted to lower masses when compared
with the core mass distribution in the Aquila Rift clouds. However, our ALMA data resolve
structures $\sim$ 4 times smaller than those achieved in the Aquila study
(Aquila is $\sim$ 9 times closer than Carina but the spatial resolution
of ALMA is 36 times better than the Herschel data used in Aquila).
The smallest mass concentrations identified with ALMA would
merge into fewer more massive structures at Herschel's resolution. 
The effect of instrumental resolution on the properties of clumps
extracted from observations has been noted before by several researchers
\citep[e.g.][for molecular observations of Carina]{schneider04}.
A new CARMA study of C$^{18}$O clumps in Orion~A taken with
a resolution of 3300~au \citep{takemura21} compares more closely
with our resolution of 2300~au, and found a relatively flat clump mass function between 
0.07M$_\odot$ and 0.7 M$_\odot$, similar to that in Fig.~\ref{fig:imf}.
Because the distribution of clump masses we find is similar to that of the IMF, the clumps do
not necessarily need to evolve further, for example by merging, to become precursors to stellar
systems. However, as we will see below, the smallest clumps tend to be unbound
gravitationally, implying that some additional processes of fragmentation and merging
will be needed to create a typical IMF.

We need to estimate the radius of each clump to find its average density and to
assess whether or not the clump is bound gravitationally. Gaussian clumps do not have a
fixed radius, but are instead characterized by their standard deviations $\sigma$ along
the major and minor axes, and a position angle.
Table~\ref{tab:clumps} gives the observed values of the
Gaussian sigmas for both the major and minor axes, integrated over the velocity extent of the
clump. We do not want to use the sizes within any single velocity slice, as those
will underestimate the true size of the clump if any rotation or otherwise
structured motion exists within the clump. After correcting for the ALMA
beam size (FWHM = 1.0$^{\prime\prime}$, $\sigma_{res}$ = 0.42$^{\prime\prime}$),
we average the Gaussian clump's sigmas along the two spatial axes to obtain 
the spatial size $<\sigma_R>$.
As most of the mass in a Gaussian clump is contained within a distance of 3-sigma of the
center, we adopt the radius of the clump to be R = 3$<\sigma_R>$.

\begin{figure}[!t]
\centering
\includegraphics[width=1.0\linewidth]{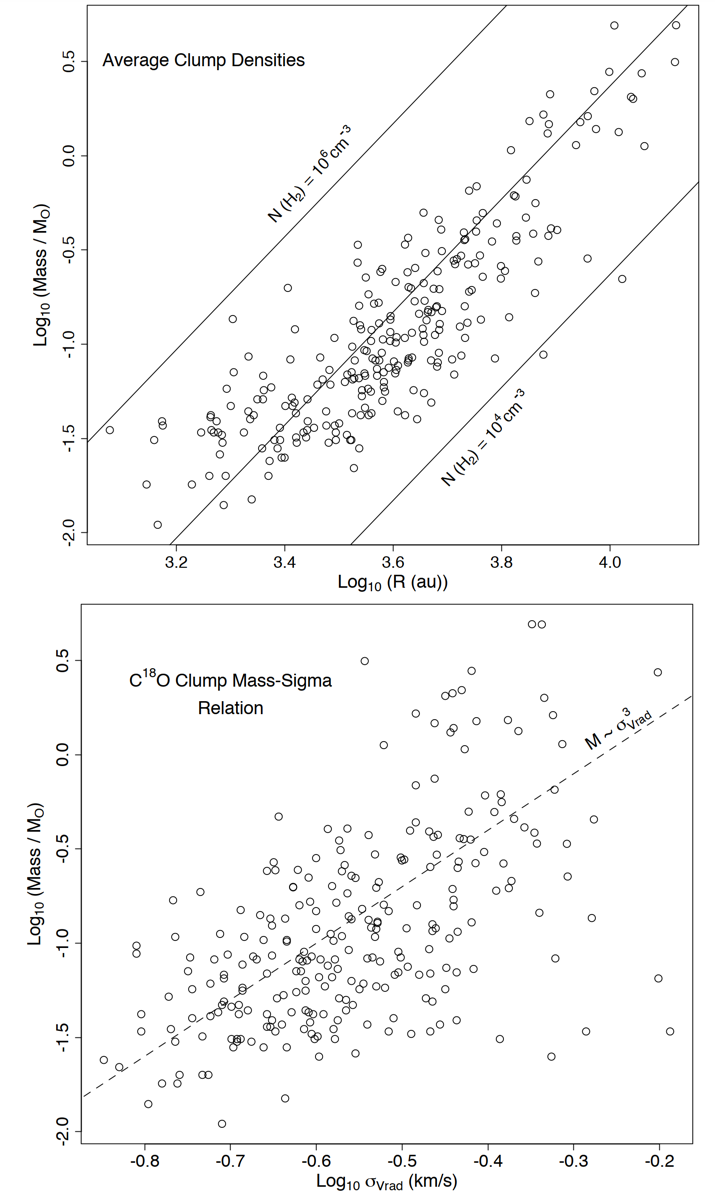}
\caption{
Top: Average densities within C$^{18}$O clumps.
Bottom: Relationship between the clump mass and the internal velocity
dispersion in the C$^{18}$O line. The dashed line is what would be
expected from a simple virial model for a constant density sphere.
}
\label{fig:densities} 
\end{figure}

The average clump density of H$_2$ is about $10^5$ cm$^{-3}$
(top panel of Fig.~\ref{fig:densities}), with a scatter of about a
factor of $\sim$ 3 on either side of this value.
The data span about an order of magnitude in radius, and have a weak trend 
in that larger clumps tend to have slightly lower average densities.
Most of the clumps are rather oblate in shape, with a median eccentricity
of 0.75 for the clumps with masses above the median mass (e.g. Fig.~\ref{fig:gaussmovie}).
The clump densities in Fig.~\ref{fig:densities} are on average
higher than those in the Aquila Rift and Orion~A cores by a factor of 
$\sim$ 3 \citep{AquilaCores,takemura21}, but these
differences are not unexpected given the larger beam sizes used for those studies.
Overall, there is nothing extraordinary about the average clump densities
in Carina's Western Wall cloud.

The C$^{18}$O clumps in Table~\ref{tab:clumps} show a clear positive correlation between their internal radial
velocity dispersions $\sigma_{\rm Vrad}$ and their Masses (bottom panel of Fig.~\ref{fig:densities}). 
For a constant density clump, M $\sim$ R$^3$, and in virial equilibrium $\sigma_{\rm Vrad}$$^2$ $\sim$ M/R. 
Hence, we expect M $\sim$ $\sigma_{\rm Vrad}^3$ for this simple model, shown as a dashed line
in the figure. This relationship fits the observations reasonably well, though the scatter is large.

\begin{figure}[!t]
\centering
\includegraphics[width=1.0\linewidth]{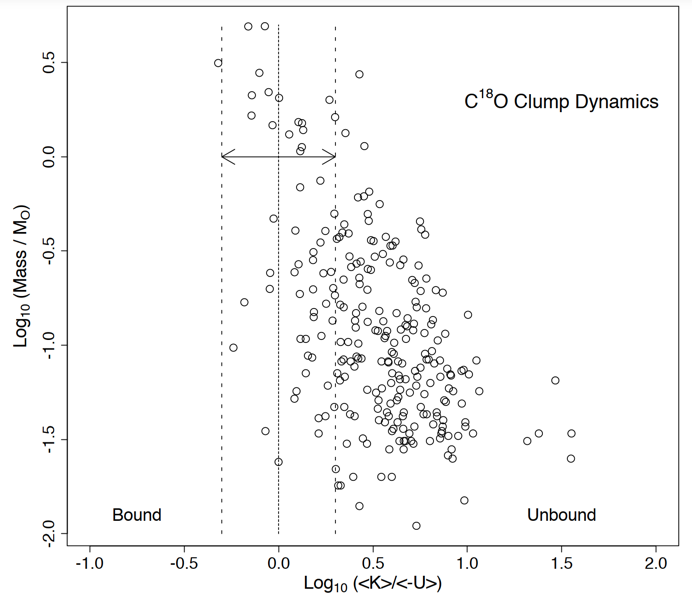}
\caption{
Ratio of the total kinetic energy <K> to the total potential energy <U>
for C$^{18}$O clumps of different masses. The boundary 
where log$_{10}$<K>/$-$<U> = 0 separates unbound clumps from bound
clumps. The other two vertical lines indicate factors of two uncertainties
about this value. Such uncertainties, can arise from differences
in how the mass is distibuted within the clumps and the radius used
to define the full extent of the clump.
}
\label{fig:dynamics} 
\end{figure}

The gravitational potential energy for a sphere of radius R
is $-\xi$GM$^2$/R, where $\xi$ is a constant of order unity that depends upon the radial
distribution of mass in the sphere. The value of $\xi$ equals 0.6 for a uniform density
sphere and increases as the density becomes more centrally concentrated. For example,
$\xi$ = 1.2 for a polytrope with polytropic index $\gamma$ = 1.4, as
appropriate for purely diatomic molecular gas. 
The kinetic energy of the clump is given by <K> = 0.5M<V$^2$>,
where <V$^2$> = 3<V$^2_{rad}$> = 3<$\sigma^2_{Vrad}$>.
Here, V$_{rad}$ is the radial velocity relative to the mean clump velocity and 
<$\sigma_{Vrad}$> is the radial velocity sigma corrected for instrumental broadening
(instrumental FWHM = 2 channels = 0.28 km$\,$s$^{-1}$,
so $\sigma_{Vinst}$ = 0.12 km$\,$s$^{-1}$). The criterion for a bound clump then 
becomes <K> $<$ <$-$U>, or M $>$ 1.5R$\sigma_{Vrad}^2$/($\xi$G). To within a
factor of order unity, this is the Bonnor-Ebert mass \citep{bonnor56}.

Fig.~\ref{fig:dynamics} shows the ratio of kinetic energy to potential energy
for the C$^{18}$O clumps in our datacube. The heavy-dashed line corresponds to $\xi$ = 1, and
the light-dashed lines on either side depict a factor of two error in the
value of <K>/<$-$U>. 
The results show that the most massive clumps congregate near the
boundary of stability, while the clumps less massive than $\sim$ 0.5 M$_\odot$
become progressively less gravitationally bound on average.

\begin{figure*}[!t]
\centering
\includegraphics[width=1.0\linewidth]{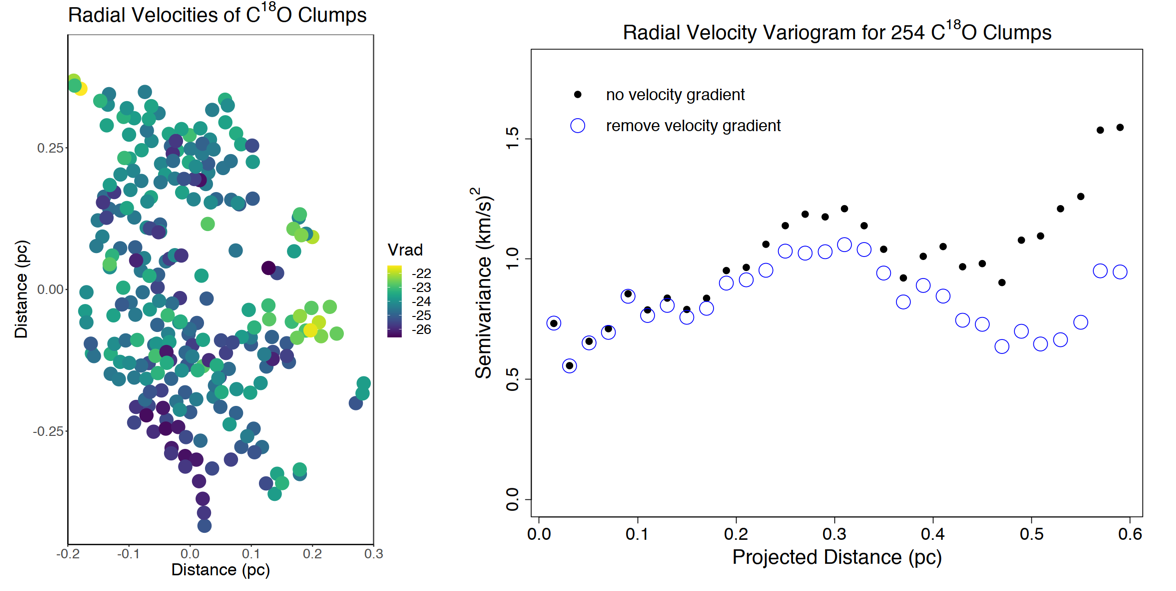}
\caption{
Left: Map of the LSR radial velocities Vrad of the C$^{18}$O clumps. The color scale
units are km$\,$s$^{-1}$.
Right: Variogram of the C$^{18}$O radial velocities, with 
and without the velocity gradient of the cloud removed (blue circles and black dots, respectively).
}
\label{fig:variogram} 
\end{figure*}

The observed line widths in the clumps are wider than expected for
thermal broadening of CO, but in most cases the
motions within the clumps do not need to be supersonic.
The first and third quartiles of the deconvolved C$^{18}$O Gaussian $\sigma_{Vrad}$
in the clumps are 0.19 km$\,$s$^{-1}$ and 0.32 km$\,$s$^{-1}$, respectively. 
The isothermal sound speed of pure H$_2$ gas at
30~K is C$_S$ = 0.42 km$\,$s$^{-1}$, implying $\sigma_{Cs}$ = C$_S$/$\gamma^{0.5}$ = 0.35
km$\,$s$^{-1}$, which is on the order of the observed motions. For reference,
the expected thermal $\sigma_{Vrad}$ for C$^{18}$O
at 30~K is only 0.09 km$\,$s$^{-1}$. 
Hence, subsonic turbulence can account for the observed
velocity broadening in most 
clumps, and any nonthermal magnetic broadening will also contribute to the
line widths.

\subsection{Kinematics Between Clumps}
\label{sec:clump-kinematics}

The left side of Fig.~\ref{fig:variogram} shows the clump V$_{LSR}$
radial velocities across the Western Wall cloud. 
The radial velocites of our collection of 254 clumps
vary between $-$26.59 km s$^{-1}$ and $-$21.46 km s$^{-1}$,
with a median of $-$24.44 km s$^{-1}$.  
There is a small velocity gradient from south to north across the cloud
(Sec.~\ref{sec:kinematics}, Fig.~\ref{fig:movie}). A good way to assess
correlation lengths of spatial data such as these is with a variogram
like the one shown on the right side of Fig.~\ref{fig:variogram}
\citep[see, e.g.,][for a description of the mathematics of variograms]{fbabu}. 
In this figure, the variogram calculates the velocity scatter within an annulus
of a given projected distance from each point. If there were a preferred
correlation length, for example from individual clouds that fragmented
into clumps, then the variogram should reveal the characteristic size
of the parent clouds.

However, when corrected for the velocity gradient across the cloud,
the variogram in Fig.~\ref{fig:variogram}
is essentially flat with projected distance, indicating that the velocity
scatter from nearby clumps is the same as that for more distant clumps. There
is a weak peak at 0.3~pc, and another one at 0.6~pc, but the latter is of low
confidence because 0.6~pc is a significant fraction of the size of the entire
mapped area.  The takeaway here is that once the velocity gradient is removed 
there are no other significant velocity concentrations present in the clump data.

\begin{figure}[t]
\includegraphics[width=1.0\linewidth]{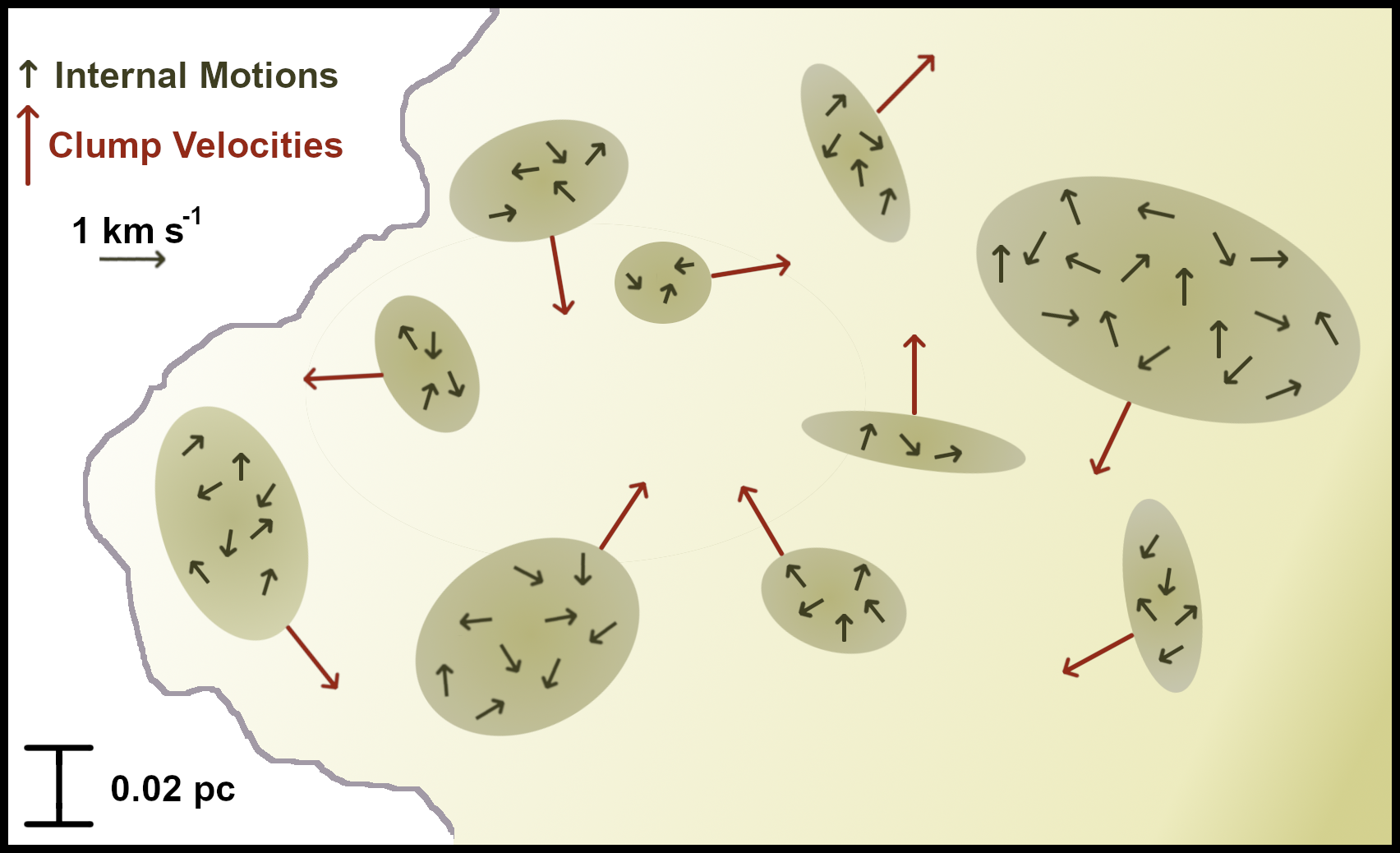}
\caption{
Summary of the observed dynamics of clumps in the Western Wall molecular cloud. 
The internal velocity dispersions within single clumps, depicted by dark arrows,
are on the order of the sound speed in the gas and are larger in the more
massive clumps. Motions between clumps, shown as red arrows,
are typically 3 $-$ 5 times larger than the sound speed.
}
\label{fig:cartoon} 
\end{figure}

The typical value of $\sim$ 0.7 km$^2$s$^{-2}$ for the semivariance implies
1.4 for the variance, or $\sim$ 1.2 km$\,$s$^{-1}$ for $\sigma_{Vrad}$. This
value is about a factor of three to five times larger than the velocity dispersion within
a typical clump.  This finding is consistent with the typical core
velocity dispersion properties detailed in the models of \citet{droplets}, who found
markedly smaller velocity dispersions within individual cores as compared with
the velocity dispersion of the cloud as a whole.
The overall dynamical picture is summarized graphically in Fig.~\ref{fig:cartoon}.

\section{Discussion}
\label{sec:discussion}

\subsection{Are There Protostars in the Western Wall?} 
\label{sec:protostars}

In Section~\ref{sec:core_prop}, we found that several of the most massive clumps in 
Western Wall cloud approach or exceed the threshold to be gravitationally bound.
This result raises the question of whether these clumps may already
have formed protostars at their centers, which would imply we should  
uncover them as strong sources of infrared continuum \citep[see, e.g.,][]{ProtostarIR}.

To investigate this possibility, we searched the standard infrared survey
catalogs for point sources over the range of the ALMA map. 2MASS \citep{2MASS} detected
three point sources within the Western Wall region, but these sources 
do not overlap with any of our clumps, and are likely to be background objects.
\citet{brooks01} observed 4.8~GHz continuum along the PDR of the Western Wall cloud, but
their maps have a 10-arcminute field of view so it is difficult to compare them with the ALMA
data. The one compact source they noted is located well to the north and west
of our map.  Of the 642 point-like Herschel continuum sources found between 70 $\mu$m and 500 $\mu$m
in Carina by \citet{gaczkowski13}, only J104331.2–593529 and J104331.8–593554
lie within our ALMA field, and neither of these coincides with a C$^{18}$O clump.
\citet[][see also \cite{tapia15}]{gaczkowski13} observed diffuse 160 $\mu$m
emission along the Western Wall, but this emission is not clearly associated with a protostar. 
The H$_2$ adaptive optics image \citep[Fig.~\ref{fig:h2c1c18};][]{hartigan20}
shows over 100 stars visible at 2.12 $\mu$m, but none of these show any
obvious relationship with the ALMA C$^{18}$O clumps.
The lack of far-IR sources in the portion of the Western Wall cloud we mapped
with ALMA implies that the clumps we identified in C$^{18}$O are starless.

In the last two
decades, several theoretical works have calculated the temperature of starless
cores as a function of physical parameters such as the gas density, the dust
opacity, and the interstellar radiation field \citep[e.g.,][]{Evans01, 
Stamatellos03, Stamatellos04}. In general, these studies found that varying
the gas density and dust opacity within reasonable ranges leads to relatively small
temperature variations. However, an increase of the external radiation intensity
from 1$G_0$ to 100$G_0$ could double the temperature at the center of a
dense core to $\gtrsim$ 20~K \citep[][]{Galli02, Lippok16}. In the
Western Wall cloud the radiation field is $\sim$ 3$\times$10$^4$ G$_0$ \citep{Wu18},
comparable to what is observed in the Orion Nebula, so the observed brightness
temperatures in Fig.~\ref{fig:co_isotope} make sense from a theoretical standpoint
without a need to invoke additional internal sources of radiation. 

\subsection{Effects of External Illumination on the Molecular Gas in the Western Wall Cloud}
\label{sec:illum}

Our new ALMA maps of the Western Wall cloud have 
shown that strong external radiation fields affect some characteristics of
star-forming regions, while leaving others relatively unchanged. The most
obvious effect of radiation occurs along the surface of the PDR, where
high-resolution images in H$_2$ uncover a variety of intriguing structures
such as waves, Kelvin-Helmholz shear, and irregular interfaces amplified
by shadowing as the gas photoablates from the surface of the cloud. 
This photoevaporation reduces the
lifetime of the molecular cloud, and, in turn, the time available for planets
to form within circumstellar disks embedded in this environment. 
Another obvious effect of the radiation is to increase the overall density
of the molecular cloud along the dissociation front, as evidenced by the
increased number of C$^{18}$O clumps situated near the surface of the cloud
compared with the number present in the interior.
In this sense, radiation helps to trigger star formation in that
it creates conditions immediately behind the front where the densities
are closer to those needed for gravitational collapse. Other less-obvious effects
of radiation include changing the isotopic CO flux
ratio plots in Fig.~\ref{fig:cotau} as discussed in Sec.~\ref{sec:opt-depth}.

On the other hand, the ALMA data do not reveal any cases where
we can clearly point to a new protostar that is forming owing to
the compression caused by the radiation front. The C$^{18}$O clumps that
congregate behind the dissociation front all appear to be starless.
Unlike in the southern
reaches of the Carina Nebula, we do not observe any pillars jutting
out of the Western Wall. There are no instances of long pillars tailing
away from the radiation front with a protostar located at the head
of the pillar where one might argue that the radiation
front helped to confine the protostar laterally after the front moved
past the collapsing core. It is possible that magnetic fields play a 
major role in inhibiting pillar development in the Western Wall, as
the morphology of the H$_2$ image along the curved portion of
the PDR immediately facing Trumpler 14 shows a series of relatively
smooth ridges \citet{hartigan20}. Higher internal temperatures for 
the Western Wall caused by the radiation field may also play a role
to inhibit star formation. Our maps do not cover the
entire cloud, and studies over a wider field of view may yet uncover
protostar activity in this region.

Because our ALMA data have extraordinarily high
spatial resolution, we were able to identify clumps that are significantly
below the mass cutoff limit of core surveys in many other regions. Nonetheless,
as for the previous core survey results, the overall shape of the mass
distribution of the C$^{18}$O clumps resembles that of the IMF, and the clump
densities and internal velocities seem fairly ordinary.
There is no clear indication that the fragmentation and merging processes
in the Western Wall differ significantly from those in less-exteme
environments, despite the fact that the temperatures are higher in this
cloud than they are in more quiescent regions.

\section{Conclusions}
\label{sec:conclusions}

The G287.38-0.62 (Car 1-E) region in the Carina star-forming 
complex provides an ideal environment to study how strong
radiation fields from massive stars influence the observable
properties of the molecular clouds in their vicinity.
Our program combined ALMA's 7-m, 12-m, and Total Power arrays to create
1$^{\prime\prime}$ (0.011~pc) resolution datacubes and continuum maps of
$^{12}$CO, $^{13}$CO, C$^{18}$O, and [C~I] over a field of view of $\sim$
60$^{\prime\prime}$ $\times$ 80$^{\prime\prime}$ (0.66 $\times$ 0.88~pc).
Situated near the young star cluster Trumpler 14, 
the Western Wall cloud is the most highly-irradiated structure in the region,
though our maps reveal several molecular clouds with different radial
velocities that superpose upon the line of sight. At least one of
these clouds is part of the Carina complex, so we can compare
the Western Wall cloud with a less-irradiated counterpart in the
same map.

In agreement with theoretical expectations, there is a clear
progression from fluoresced H$_2$, to [C~I], to C$^{18}$O
with distance into the PDR front of the Western Wall, with spatial
offsets between these regions $\sim$ $10^{17}$ cm.
Emission from the optically thick $^{12}$CO line extends across
the region, while $^{13}$CO, [C~I] and especially C$^{18}$O are more optically thin, 
and concentrate into clumps and filaments closer to the PDR interface.

The temperature in the Western Wall cloud reaches $\sim$ 80~K in its
outer layes, about 20\%\ higher than that within the less-irradiated cloud.
As predicted by theoretical models, C$^{18}$O must
be depleted relative to $^{13}$CO to explain the observed flux ratios
among the CO isotopologues in both clouds.
The flux ratios are generally similar between the two clouds, though
some systematic differences exist.
Dust sizes inferred from continuum measurements at two frequencies
are consistent with interstellar grains, with no evidence for grain growth
in the Western Wall region.  We discovered several bright, compact sources in the CO cubes
that may represent disks or outflows from young stars, but these all have velocities
significantly different from those in the Western Wall
cloud, and so are most likely to be foreground or background objects.

Using a Gaussian fitting algorithm, we identified 254
distinct C$^{18}$O clumps in the Western Wall cloud. The
mass distribution of these objects is similar to that of the stellar IMF. 
More massive clumps exhibit higher internal velocity dispersions
than less massive clumps do, and the line widths generally follow what is
expected from virial equilibrium, with significant scatter.
Smaller clumps tend to be weakly unbound, and more massive clumps
are more likely to be bound. The average clump density is $\sim$ $10^5$
cm$^{-3}$ across clumps of all sizes. A typical clump radius is 0.02~pc,
or $\sim$ 4 ALMA beam sizes. Clumps tend to be oblate, with
a median eccentricity of 0.75 for clumps with masses
above the median mass. 

The observed line widths in clumps significantly exceed those expected for
simple thermal broadening in CO, but can be explained by $\sim$ Mach~1 motions
within the clumps in most cases. The variation of velocities between
clumps is higher, implying Mach numbers of 3 $-$ 5 for these relative motions.
A variogram analysis of the radial velocities of the clumps does
not show any characteristic coherent spatial scales
among the cores, though there is a velocity gradient of
a few km$\,$s$^{-1}$ north to south across the Western Wall cloud.
The ALMA data show no direct evidence for                     
triggering in the Western Wall in that the clumps and cores within the
mapped area appear starless, and no pillars are present. However,
the densest portion of the cloud lies closest to the PDR,
and some of the C$^{18}$O emission is flattened along the radiation front.

The extraordinary spatial resolution of ALMA as compared with single dish
observations makes it possible to compare molecular maps on the same
footing as optical and near-infrared images. This capability shows its
true power in a complex region like Carina, where several molecular clouds 
exist along the line of sight, intricate interface shapes abound, and hundreds of stars
are visible within a square parsec. The high spatial resolution of ALMA
enables studies of line ratios and kinematics like those presented in this paper
that explore scales ranging from a few thousand au to $\sim$ 1 pc
in a single cohesive data set. This
multi-scale observational capability is an excellent way to inform analogous
models as to the main physical properties at work in these environments.
 
\clearpage
\startlongtable
\begin{deluxetable*}{cccccccccc}
\tabletypesize{\scriptsize}
\tablecaption{Gaussian model fits for C$^{18}$O clumps.
Column 1: ID number of the clump; Columns 2 and 3: Coordinates;
Column 4: V$_{LSR}$ of the clump; Columns 5 and 6: sigma of the Gaussian clump
shape in arcseconds along major and minor axes respectively, summed over velocity and
corrected for the 0.47$^{\prime\prime}$ 1-sigma ALMA
beam size; Column 7: Position angle of the Gaussian clump's major
axis on the sky when integrated over velocity;
Column 8: Sigma of Gaussian clump dispersion in km/s integrated over
the spatial extent of the clump after correcting for
the spectral resolution of the observations; Column 9: Peak flux in Jy beam$^{-1}$;
Column 10: Mass in solar units.
}
\label{tab:clumps}
\tablehead{\\
\colhead{ID} & \colhead{RA (2000.0)} & \colhead{Dec (2000.0)} & \colhead{V$_{LSR}$} 
 & \colhead{dMax} & \colhead{dMin} &  \colhead{PA} & \colhead{Width} & \colhead{Flux} &\colhead{Mass}\\
 &  &  & (km$\,$s$^{-1}$) & (arcsec) & (arcsec)
 & (deg) & (km s$^{-1}$) & (Jky/beam) & (M$_\odot$)
}
\colnumbers
\startdata
&\\
1 & 10:43:31.64 & $-$59:35:38.754 & $-$24.804 & 2.05 & 1.04 &  88 & 0.46 & 0.93 & 4.92 \\ 
2 & 10:43:30.71 & $-$59:35:57.119 & $-$24.927 & 2.77 & 1.18 & 118 & 0.45 & 0.74 & 4.93 \\ 
3 & 10:43:30.71 & $-$59:35:51.174 & $-$25.523 & 1.50 & 0.92 & 141 & 0.36 & 0.70 & 2.12 \\ 
4 & 10:43:31.77 & $-$59:35:43.146 & $-$24.348 & 1.45 & 0.90 &  92 & 0.33 & 0.64 & 1.66 \\ 
5 & 10:43:30.11 & $-$59:35:58.608 & $-$25.059 & 1.67 & 1.34 & 168 & 0.38 & 0.56 & 2.79 \\ 
6 & 10:43:30.78 & $-$59:36:14.024 & $-$25.920 & 1.80 & 1.05 & 158 & 0.37 & 0.54 & 2.21 \\ 
7 & 10:43:30.82 & $-$59:35:28.015 & $-$24.123 & 2.65 & 1.28 &  85 & 0.29 & 0.52 & 3.15 \\ 
8 & 10:43:31.03 & $-$59:35:48.537 & $-$24.524 & 1.22 & 1.01 &  90 & 0.42 & 0.49 & 1.53 \\ 
9 & 10:43:30.88 & $-$59:36:09.897 & $-$25.436 & 1.31 & 1.09 &  24 & 0.34 & 0.51 & 1.47 \\ 
10 & 10:43:30.18 & $-$59:36:07.792 & $-$25.092 & 2.07 & 1.22 & 169 & 0.35 & 0.47 & 2.05 \\ 
11 & 10:43:31.23 & $-$59:35:35.765 & $-$24.134 & 1.22 & 0.87 &  44 & 0.37 & 0.44 & 1.07 \\ 
12 & 10:43:31.25 & $-$59:35:30.799 & $-$23.608 & 0.97 & 0.88 &   2 & 0.33 & 0.43 & 0.69 \\ 
13 & 10:43:31.23 & $-$59:35:53.561 & $-$24.749 & 1.43 & 0.95 &  67 & 0.36 & 0.42 & 1.31 \\ 
14 & 10:43:31.43 & $-$59:35:44.995 & $-$24.833 & 1.96 & 0.84 &  85 & 0.47 & 0.41 & 1.62 \\ 
15 & 10:43:30.70 & $-$59:35:49.172 & $-$24.980 & 1.92 & 0.96 & 135 & 0.36 & 0.41 & 1.39 \\ 
16 & 10:43:30.88 & $-$59:36:03.166 & $-$25.354 & 0.94 & 0.63 &  21 & 0.38 & 0.38 & 0.50 \\ 
17 & 10:43:31.74 & $-$59:35:36.834 & $-$24.763 & 0.76 & 0.44 & 155 & 0.24 & 0.40 & 0.20 \\ 
18 & 10:43:31.16 & $-$59:36:05.457 & $-$24.521 & 1.70 & 1.00 & 112 & 0.39 & 0.37 & 1.51 \\ 
19 & 10:43:31.61 & $-$59:36:01.706 & $-$23.382 & 1.02 & 0.65 & 141 & 0.27 & 0.35 & 0.41 \\ 
20 & 10:43:31.16 & $-$59:36:12.577 & $-$25.519 & 0.80 & 0.60 &  93 & 0.22 & 0.34 & 0.24 \\ 
21 & 10:43:30.77 & $-$59:36:00.291 & $-$24.838 & 2.50 & 0.85 & 156 & 0.46 & 0.34 & 2.01 \\ 
22 & 10:43:31.16 & $-$59:35:40.116 & $-$24.508 & 2.05 & 1.38 &  84 & 0.63 & 0.34 & 2.74 \\ 
23 & 10:43:31.86 & $-$59:35:40.580 & $-$24.478 & 0.87 & 0.49 &  81 & 0.37 & 0.33 & 0.27 \\ 
24 & 10:43:31.63 & $-$59:35:47.997 & $-$24.044 & 1.05 & 0.76 &   8 & 0.48 & 0.32 & 0.65 \\ 
25 & 10:43:30.64 & $-$59:35:31.627 & $-$24.236 & 2.57 & 0.93 &  61 & 0.30 & 0.32 & 1.13 \\ 
26 & 10:43:29.40 & $-$59:35:21.485 & $-$23.355 & 0.78 & 0.72 & 177 & 0.34 & 0.30 & 0.37 \\ 
27 & 10:43:30.34 & $-$59:36:09.265 & $-$24.607 & 0.96 & 0.70 & 177 & 0.27 & 0.30 & 0.31 \\ 
28 & 10:43:31.14 & $-$59:35:55.510 & $-$24.524 & 0.78 & 0.54 &  79 & 0.49 & 0.30 & 0.34 \\ 
29 & 10:43:30.03 & $-$59:36:00.336 & $-$25.677 & 1.60 & 0.65 & 149 & 0.23 & 0.29 & 0.47 \\ 
30 & 10:43:29.73 & $-$59:35:33.006 & $-$24.309 & 1.20 & 1.01 & 134 & 0.34 & 0.29 & 0.75 \\ 
31 & 10:43:30.28 & $-$59:35:52.879 & $-$25.736 & 1.03 & 0.63 &   8 & 0.43 & 0.29 & 0.46 \\ 
32 & 10:43:30.54 & $-$59:36:12.143 & $-$25.451 & 1.03 & 0.62 &  27 & 0.23 & 0.29 & 0.24 \\ 
33 & 10:43:30.69 & $-$59:35:23.617 & $-$24.376 & 1.37 & 0.76 & 108 & 0.41 & 0.29 & 0.62 \\ 
34 & 10:43:30.55 & $-$59:36:13.531 & $-$26.438 & 1.02 & 0.54 & 145 & 0.17 & 0.28 & 0.17 \\ 
35 & 10:43:30.56 & $-$59:35:33.370 & $-$23.815 & 0.87 & 0.63 &  41 & 0.27 & 0.26 & 0.24 \\ 
36 & 10:43:29.40 & $-$59:35:25.062 & $-$23.623 & 1.00 & 0.78 &  58 & 0.37 & 0.26 & 0.36 \\ 
37 & 10:43:30.97 & $-$59:35:46.556 & $-$24.298 & 0.91 & 0.82 &  78 & 0.25 & 0.26 & 0.28 \\ 
38 & 10:43:30.14 & $-$59:36:03.509 & $-$25.191 & 2.21 & 0.94 &  17 & 0.43 & 0.26 & 1.34 \\ 
39 & 10:43:30.50 & $-$59:35:58.547 & $-$24.326 & 1.10 & 0.75 & 134 & 0.32 & 0.26 & 0.40 \\ 
40 & 10:43:31.02 & $-$59:35:29.477 & $-$23.656 & 0.87 & 0.48 &  71 & 0.15 & 0.25 & 0.10 \\ 
41 & 10:43:30.63 & $-$59:36:07.488 & $-$25.579 & 1.19 & 0.94 &   2 & 0.29 & 0.25 & 0.38 \\ 
\\
42 & 10:43:30.44 & $-$59:36:16.615 & $-$26.216 & 0.74 & 0.62 & 148 & 0.27 & 0.24 & 0.16 \\ 
43 & 10:43:26.88 & $-$59:36:09.501 & $-$25.230 & 0.74 & 0.52 & 176 & 0.21 & 0.24 & 0.11 \\ 
44 & 10:43:29.80 & $-$59:36:28.907 & $-$25.243 & 1.08 & 0.71 &  56 & 0.34 & 0.24 & 0.39 \\ 
45 & 10:43:31.18 & $-$59:35:32.735 & $-$24.230 & 1.25 & 0.63 &   9 & 0.22 & 0.24 & 0.27 \\ 
46 & 10:43:31.43 & $-$59:36:02.974 & $-$23.853 & 1.00 & 0.54 &  25 & 0.26 & 0.23 & 0.20 \\ 
47 & 10:43:31.12 & $-$59:36:10.105 & $-$25.803 & 1.37 & 0.47 & 113 & 0.39 & 0.24 & 0.27 \\ 
48 & 10:43:30.55 & $-$59:35:47.066 & $-$24.616 & 1.33 & 0.81 & 146 & 0.40 & 0.24 & 0.61 \\ 
49 & 10:43:30.43 & $-$59:35:53.745 & $-$24.733 & 1.69 & 0.97 & 163 & 0.49 & 0.23 & 1.14 \\ 
50 & 10:43:30.49 & $-$59:35:26.901 & $-$23.579 & 1.29 & 0.69 &  70 & 0.27 & 0.23 & 0.35 \\ 
51 & 10:43:30.91 & $-$59:35:26.376 & $-$24.594 & 0.87 & 0.65 & 143 & 0.24 & 0.23 & 0.20 \\ 
52 & 10:43:31.16 & $-$59:35:24.412 & $-$23.747 & 1.03 & 0.86 & 170 & 0.41 & 0.23 & 0.50 \\ 
53 & 10:43:29.57 & $-$59:35:37.227 & $-$24.905 & 0.81 & 0.55 & 168 & 0.27 & 0.22 & 0.18 \\ 
54 & 10:43:31.98 & $-$59:36:01.642 & $-$23.630 & 0.78 & 0.62 &  91 & 0.37 & 0.22 & 0.25 \\ 
55 & 10:43:30.99 & $-$59:35:51.958 & $-$24.712 & 0.86 & 0.49 & 134 & 0.30 & 0.22 & 0.16 \\ 
56 & 10:43:31.69 & $-$59:35:40.166 & $-$25.586 & 0.76 & 0.41 & 141 & 0.22 & 0.22 & 0.09 \\ 
57 & 10:43:30.84 & $-$59:35:36.921 & $-$23.568 & 0.89 & 0.60 &   4 & 0.45 & 0.22 & 0.34 \\ 
58 & 10:43:31.61 & $-$59:36:04.859 & $-$24.526 & 0.78 & 0.60 & 175 & 0.25 & 0.21 & 0.17 \\ 
59 & 10:43:32.33 & $-$59:35:18.473 & $-$22.183 & 0.98 & 0.61 & 108 & 0.39 & 0.21 & 0.30 \\ 
60 & 10:43:31.42 & $-$59:35:39.024 & $-$24.743 & 0.59 & 0.46 &  85 & 0.53 & 0.21 & 0.14 \\ 
61 & 10:43:29.63 & $-$59:36:08.484 & $-$24.415 & 0.94 & 0.84 &  92 & 0.37 & 0.20 & 0.36 \\ 
62 & 10:43:31.54 & $-$59:36:00.121 & $-$24.095 & 0.83 & 0.70 & 170 & 0.34 & 0.20 & 0.25 \\ 
63 & 10:43:29.51 & $-$59:36:05.349 & $-$25.600 & 1.18 & 0.81 & 128 & 0.33 & 0.20 & 0.44 \\ 
64 & 10:43:30.17 & $-$59:36:17.875 & $-$26.385 & 0.78 & 0.65 &  89 & 0.22 & 0.20 & 0.14 \\ 
65 & 10:43:27.55 & $-$59:35:58.912 & $-$22.465 & 1.07 & 0.81 &  94 & 0.29 & 0.20 & 0.30 \\ 
66 & 10:43:29.65 & $-$59:35:23.098 & $-$24.397 & 1.06 & 0.67 &  53 & 0.32 & 0.19 & 0.28 \\ 
67 & 10:43:31.14 & $-$59:35:44.848 & $-$25.034 & 1.25 & 0.63 &  75 & 0.53 & 0.20 & 0.45 \\ 
68 & 10:43:30.09 & $-$59:35:27.107 & $-$23.026 & 1.04 & 0.56 &   9 & 0.30 & 0.18 & 0.21 \\ 
69 & 10:43:28.90 & $-$59:36:00.208 & $-$25.128 & 1.64 & 0.68 & 154 & 0.41 & 0.17 & 0.56 \\ 
70 & 10:43:30.92 & $-$59:36:05.661 & $-$24.037 & 0.89 & 0.56 &  69 & 0.23 & 0.17 & 0.14 \\ 
71 & 10:43:30.95 & $-$59:36:08.944 & $-$24.809 & 0.95 & 0.39 &  99 & 0.25 & 0.18 & 0.09 \\ 
72 & 10:43:27.84 & $-$59:35:58.018 & $-$22.936 & 0.85 & 0.77 &  88 & 0.29 & 0.18 & 0.20 \\ 
73 & 10:43:31.89 & $-$59:35:44.640 & $-$24.468 & 0.69 & 0.43 & 131 & 0.24 & 0.17 & 0.07 \\ 
74 & 10:43:31.56 & $-$59:35:55.610 & $-$23.623 & 0.91 & 0.48 &  56 & 0.49 & 0.17 & 0.23 \\ 
75 & 10:43:31.03 & $-$59:36:03.548 & $-$24.768 & 1.30 & 0.96 &  59 & 0.45 & 0.17 & 0.39 \\ 
76 & 10:43:31.67 & $-$59:35:22.272 & $-$24.066 & 0.92 & 0.84 &  90 & 0.35 & 0.17 & 0.29 \\ 
77 & 10:43:29.83 & $-$59:36:03.646 & $-$22.839 & 0.77 & 0.54 &   7 & 0.29 & 0.17 & 0.13 \\ 
78 & 10:43:28.07 & $-$59:35:45.473 & $-$23.748 & 0.90 & 0.75 & 168 & 0.20 & 0.17 & 0.15 \\ 
79 & 10:43:31.54 & $-$59:35:36.100 & $-$25.732 & 0.79 & 0.44 &  50 & 0.32 & 0.17 & 0.12 \\ 
80 & 10:43:31.37 & $-$59:35:51.233 & $-$23.449 & 0.75 & 0.69 &  85 & 0.42 & 0.16 & 0.21 \\ 
81 & 10:43:31.26 & $-$59:35:27.015 & $-$23.767 & 0.79 & 0.57 &  29 & 0.23 & 0.16 & 0.10 \\ 
82 & 10:43:31.32 & $-$59:35:59.581 & $-$24.764 & 1.43 & 0.62 & 130 & 0.25 & 0.16 & 0.22 \\ 
83 & 10:43:29.96 & $-$59:36:08.892 & $-$24.451 & 0.79 & 0.40 &  17 & 0.18 & 0.16 & 0.06 \\ 
84 & 10:43:30.88 & $-$59:35:24.546 & $-$23.601 & 0.73 & 0.45 &  91 & 0.17 & 0.16 & 0.05 \\ 
85 & 10:43:29.95 & $-$59:35:56.818 & $-$25.117 & 0.55 & 0.38 &  43 & 0.17 & 0.16 & 0.04 \\ 
86 & 10:43:30.86 & $-$59:35:54.685 & $-$23.835 & 1.38 & 1.09 &  93 & 0.26 & 0.15 & 0.40 \\ 
87 & 10:43:29.33 & $-$59:36:04.085 & $-$24.782 & 1.01 & 0.88 &  83 & 0.28 & 0.15 & 0.23 \\ 
88 & 10:43:31.63 & $-$59:35:34.984 & $-$23.753 & 1.87 & 0.51 &  49 & 0.18 & 0.16 & 0.19 \\ 
89 & 10:43:29.60 & $-$59:36:06.708 & $-$24.683 & 1.72 & 0.63 & 172 & 0.32 & 0.15 & 0.28 \\ 
90 & 10:43:30.44 & $-$59:36:05.644 & $-$24.924 & 1.64 & 0.40 &  12 & 0.30 & 0.15 & 0.13 \\ 
91 & 10:43:30.33 & $-$59:35:29.511 & $-$23.474 & 0.89 & 0.74 &  78 & 0.24 & 0.15 & 0.16 \\ 
92 & 10:43:29.88 & $-$59:35:26.015 & $-$23.809 & 1.03 & 0.77 & 177 & 0.41 & 0.14 & 0.27 \\ 
\\
93 & 10:43:30.29 & $-$59:35:56.709 & $-$24.015 & 1.30 & 0.76 & 111 & 0.24 & 0.15 & 0.24 \\ 
94 & 10:43:29.91 & $-$59:36:21.853 & $-$26.078 & 0.97 & 0.63 &  88 & 0.25 & 0.15 & 0.15 \\ 
95 & 10:43:30.25 & $-$59:35:26.132 & $-$23.636 & 0.76 & 0.56 &   0 & 0.34 & 0.14 & 0.13 \\ 
96 & 10:43:27.72 & $-$59:35:43.213 & $-$22.039 & 0.94 & 0.52 & 145 & 0.22 & 0.14 & 0.10 \\ 
97 & 10:43:28.69 & $-$59:36:16.452 & $-$24.781 & 1.17 & 0.85 & 134 & 0.27 & 0.14 & 0.26 \\ 
98 & 10:43:30.07 & $-$59:35:29.280 & $-$24.453 & 1.19 & 0.94 &  92 & 0.38 & 0.14 & 0.35 \\ 
99 & 10:43:31.39 & $-$59:35:53.887 & $-$25.135 & 0.92 & 0.47 & 109 & 0.34 & 0.14 & 0.09 \\ 
100 & 10:43:27.98 & $-$59:35:40.129 & $-$24.253 & 0.77 & 0.53 & 143 & 0.18 & 0.14 & 0.07 \\ 
101 & 10:43:29.44 & $-$59:35:32.294 & $-$24.602 & 0.75 & 0.57 &  44 & 0.35 & 0.14 & 0.12 \\ 
102 & 10:43:29.77 & $-$59:35:34.826 & $-$24.429 & 1.07 & 0.58 &   4 & 0.19 & 0.14 & 0.11 \\ 
103 & 10:43:29.89 & $-$59:35:34.208 & $-$26.379 & 0.79 & 0.57 &   1 & 0.29 & 0.13 & 0.12 \\ 
104 & 10:43:30.65 & $-$59:35:34.548 & $-$24.563 & 1.07 & 0.45 &   3 & 0.34 & 0.14 & 0.12 \\ 
105 & 10:43:31.02 & $-$59:35:34.340 & $-$24.439 & 0.94 & 0.39 & 155 & 0.28 & 0.13 & 0.06 \\ 
106 & 10:43:31.36 & $-$59:35:24.205 & $-$23.328 & 1.04 & 0.57 & 100 & 0.28 & 0.13 & 0.15 \\ 
107 & 10:43:29.35 & $-$59:35:22.390 & $-$24.099 & 0.86 & 0.60 &  24 & 0.23 & 0.13 & 0.10 \\ 
108 & 10:43:31.46 & $-$59:36:05.723 & $-$24.447 & 0.66 & 0.57 &   3 & 0.19 & 0.13 & 0.06 \\ 
109 & 10:43:28.21 & $-$59:36:00.033 & $-$25.138 & 1.20 & 0.48 &  10 & 0.30 & 0.13 & 0.15 \\ 
110 & 10:43:29.19 & $-$59:35:26.830 & $-$23.425 & 0.72 & 0.58 &  60 & 0.20 & 0.12 & 0.07 \\ 
111 & 10:43:29.48 & $-$59:35:59.571 & $-$25.406 & 0.97 & 0.62 &  50 & 0.36 & 0.12 & 0.17 \\ 
112 & 10:43:27.59 & $-$59:35:56.712 & $-$21.962 & 0.74 & 0.64 & 137 & 0.38 & 0.12 & 0.13 \\ 
113 & 10:43:30.66 & $-$59:35:41.234 & $-$24.939 & 0.71 & 0.39 &  22 & 0.30 & 0.13 & 0.06 \\ 
114 & 10:43:29.72 & $-$59:36:02.703 & $-$25.934 & 0.85 & 0.54 & 134 & 0.21 & 0.12 & 0.08 \\ 
115 & 10:43:31.76 & $-$59:35:37.737 & $-$25.784 & 0.68 & 0.38 &  61 & 0.19 & 0.13 & 0.04 \\ 
116 & 10:43:31.10 & $-$59:35:59.506 & $-$23.285 & 1.54 & 0.87 & 129 & 0.35 & 0.12 & 0.38 \\ 
117 & 10:43:30.73 & $-$59:36:00.098 & $-$23.902 & 1.30 & 0.55 & 119 & 0.17 & 0.12 & 0.11 \\ 
118 & 10:43:30.86 & $-$59:35:49.359 & $-$23.446 & 0.84 & 0.61 & 177 & 0.27 & 0.12 & 0.11 \\ 
119 & 10:43:29.18 & $-$59:36:07.288 & $-$23.954 & 1.30 & 0.62 &  18 & 0.22 & 0.12 & 0.14 \\ 
120 & 10:43:31.30 & $-$59:35:22.824 & $-$23.952 & 0.88 & 0.69 &  22 & 0.26 & 0.12 & 0.11 \\ 
121 & 10:43:28.85 & $-$59:36:13.545 & $-$24.950 & 0.81 & 0.77 &  88 & 0.28 & 0.12 & 0.13 \\ 
122 & 10:43:29.57 & $-$59:36:03.513 & $-$23.458 & 1.26 & 0.59 &  51 & 0.41 & 0.12 & 0.19 \\ 
123 & 10:43:28.08 & $-$59:35:41.920 & $-$22.608 & 0.85 & 0.59 & 143 & 0.24 & 0.12 & 0.08 \\ 
124 & 10:43:27.38 & $-$59:35:54.264 & $-$22.623 & 0.96 & 0.65 & 114 & 0.19 & 0.11 & 0.08 \\ 
125 & 10:43:30.17 & $-$59:36:01.501 & $-$24.857 & 1.07 & 0.44 & 175 & 0.20 & 0.11 & 0.07 \\ 
126 & 10:43:31.94 & $-$59:36:02.057 & $-$24.653 & 0.74 & 0.57 &  82 & 0.25 & 0.11 & 0.08 \\ 
127 & 10:43:27.96 & $-$59:36:20.728 & $-$24.595 & 0.98 & 0.66 &   9 & 0.36 & 0.11 & 0.16 \\ 
128 & 10:43:29.29 & $-$59:35:31.207 & $-$24.611 & 0.54 & 0.47 &   4 & 0.22 & 0.11 & 0.04 \\ 
129 & 10:43:29.09 & $-$59:35:28.536 & $-$23.599 & 0.72 & 0.42 & 172 & 0.20 & 0.11 & 0.05 \\ 
130 & 10:43:30.09 & $-$59:35:57.785 & $-$24.770 & 0.69 & 0.37 & 116 & 0.52 & 0.11 & 0.03 \\ 
131 & 10:43:29.73 & $-$59:35:31.599 & $-$25.088 & 0.82 & 0.64 & 111 & 0.21 & 0.11 & 0.08 \\ 
132 & 10:43:31.65 & $-$59:35:20.603 & $-$24.460 & 0.66 & 0.45 &  52 & 0.33 & 0.11 & 0.07 \\ 
133 & 10:43:30.23 & $-$59:35:36.145 & $-$23.644 & 1.00 & 0.66 & 155 & 0.42 & 0.11 & 0.20 \\ 
134 & 10:43:30.85 & $-$59:36:07.713 & $-$25.265 & 0.60 & 0.42 &  71 & 0.26 & 0.11 & 0.03 \\ 
135 & 10:43:30.75 & $-$59:36:02.023 & $-$23.009 & 0.86 & 0.51 & 152 & 0.27 & 0.11 & 0.09 \\ 
136 & 10:43:27.74 & $-$59:35:54.436 & $-$22.761 & 0.95 & 0.70 &  83 & 0.25 & 0.11 & 0.12 \\ 
137 & 10:43:30.01 & $-$59:35:37.248 & $-$24.254 & 0.96 & 0.80 &  81 & 0.22 & 0.10 & 0.12 \\ 
138 & 10:43:30.90 & $-$59:35:37.596 & $-$24.136 & 0.61 & 0.40 & 134 & 0.27 & 0.10 & 0.04 \\ 
139 & 10:43:28.72 & $-$59:36:06.324 & $-$23.923 & 0.78 & 0.62 &  81 & 0.24 & 0.10 & 0.07 \\ 
140 & 10:43:30.08 & $-$59:36:10.929 & $-$25.120 & 0.85 & 0.71 &  44 & 0.29 & 0.10 & 0.12 \\ 
141 & 10:43:28.87 & $-$59:35:37.141 & $-$25.102 & 0.93 & 0.62 &  22 & 0.46 & 0.10 & 0.14 \\ 
142 & 10:43:29.49 & $-$59:36:10.097 & $-$24.959 & 1.15 & 0.65 &   3 & 0.33 & 0.10 & 0.16 \\ 
143 & 10:43:29.84 & $-$59:36:24.636 & $-$26.091 & 0.85 & 0.50 &  93 & 0.25 & 0.10 & 0.07 \\ 
\\
144 & 10:43:31.42 & $-$59:35:33.319 & $-$24.091 & 0.66 & 0.49 & 115 & 0.20 & 0.10 & 0.05 \\ 
145 & 10:43:30.48 & $-$59:35:46.647 & $-$25.645 & 0.86 & 0.66 & 173 & 0.21 & 0.10 & 0.09 \\ 
146 & 10:43:31.30 & $-$59:35:38.633 & $-$23.573 & 1.24 & 0.41 &  32 & 0.27 & 0.10 & 0.07 \\ 
147 & 10:43:30.46 & $-$59:36:02.704 & $-$25.793 & 0.72 & 0.48 &  81 & 0.22 & 0.10 & 0.04 \\ 
148 & 10:43:29.76 & $-$59:35:52.963 & $-$25.097 & 1.58 & 0.85 & 140 & 0.44 & 0.10 & 0.41 \\ 
149 & 10:43:26.75 & $-$59:36:07.963 & $-$23.755 & 0.86 & 0.45 &  30 & 0.16 & 0.10 & 0.03 \\ 
150 & 10:43:28.48 & $-$59:36:01.407 & $-$25.289 & 1.17 & 0.53 & 135 & 0.30 & 0.09 & 0.13 \\ 
151 & 10:43:28.50 & $-$59:35:59.033 & $-$24.969 & 0.57 & 0.49 &   8 & 0.26 & 0.09 & 0.04 \\ 
152 & 10:43:30.60 & $-$59:36:10.255 & $-$26.273 & 0.78 & 0.61 &  21 & 0.26 & 0.10 & 0.08 \\ 
153 & 10:43:29.53 & $-$59:36:05.532 & $-$24.104 & 0.78 & 0.64 & 114 & 0.32 & 0.09 & 0.07 \\ 
154 & 10:43:29.93 & $-$59:36:04.300 & $-$23.332 & 0.62 & 0.47 & 147 & 0.27 & 0.09 & 0.05 \\ 
155 & 10:43:29.39 & $-$59:36:01.568 & $-$25.595 & 1.08 & 0.75 &  51 & 0.36 & 0.09 & 0.19 \\ 
156 & 10:43:26.73 & $-$59:36:06.366 & $-$23.746 & 0.66 & 0.50 & 155 & 0.18 & 0.09 & 0.03 \\ 
157 & 10:43:27.25 & $-$59:35:58.501 & $-$22.659 & 1.11 & 0.66 &  82 & 0.20 & 0.09 & 0.09 \\ 
158 & 10:43:30.92 & $-$59:35:41.714 & $-$24.115 & 0.79 & 0.42 & 147 & 0.26 & 0.09 & 0.06 \\ 
159 & 10:43:30.37 & $-$59:35:46.092 & $-$24.109 & 1.06 & 0.54 & 132 & 0.26 & 0.09 & 0.10 \\ 
160 & 10:43:28.40 & $-$59:36:20.680 & $-$23.602 & 0.76 & 0.63 &  91 & 0.36 & 0.09 & 0.11 \\ 
161 & 10:43:31.12 & $-$59:35:46.912 & $-$26.098 & 0.87 & 0.62 & 134 & 0.29 & 0.09 & 0.11 \\ 
162 & 10:43:28.61 & $-$59:36:03.709 & $-$24.924 & 0.74 & 0.61 &  95 & 0.32 & 0.09 & 0.08 \\ 
163 & 10:43:29.28 & $-$59:35:37.324 & $-$24.779 & 0.71 & 0.46 & 158 & 0.48 & 0.09 & 0.08 \\ 
164 & 10:43:28.84 & $-$59:36:17.273 & $-$25.266 & 0.75 & 0.45 & 143 & 0.19 & 0.09 & 0.04 \\ 
165 & 10:43:27.96 & $-$59:35:39.623 & $-$22.969 & 0.83 & 0.67 & 149 & 0.24 & 0.09 & 0.08 \\ 
166 & 10:43:29.81 & $-$59:36:26.860 & $-$25.771 & 0.64 & 0.45 &  53 & 0.24 & 0.09 & 0.04 \\ 
167 & 10:43:29.85 & $-$59:35:30.037 & $-$24.461 & 0.91 & 0.45 &  64 & 0.24 & 0.09 & 0.06 \\ 
168 & 10:43:27.96 & $-$59:36:19.998 & $-$23.457 & 0.78 & 0.62 &  44 & 0.31 & 0.09 & 0.09 \\ 
169 & 10:43:28.87 & $-$59:35:31.328 & $-$23.801 & 0.81 & 0.51 &  61 & 0.26 & 0.09 & 0.07 \\ 
170 & 10:43:29.70 & $-$59:35:27.797 & $-$24.126 & 0.90 & 0.62 & 176 & 0.37 & 0.08 & 0.12 \\ 
171 & 10:43:28.40 & $-$59:35:48.919 & $-$25.436 & 0.68 & 0.51 & 145 & 0.23 & 0.09 & 0.05 \\ 
172 & 10:43:27.75 & $-$59:35:57.981 & $-$21.633 & 0.75 & 0.59 & 134 & 0.21 & 0.09 & 0.06 \\ 
173 & 10:43:31.35 & $-$59:35:30.704 & $-$23.127 & 0.62 & 0.45 &  76 & 0.34 & 0.08 & 0.03 \\ 
174 & 10:43:30.31 & $-$59:36:13.278 & $-$26.250 & 0.79 & 0.62 & 170 & 0.29 & 0.08 & 0.06 \\ 
175 & 10:43:28.56 & $-$59:35:48.111 & $-$26.586 & 0.50 & 0.46 &  44 & 0.35 & 0.08 & 0.04 \\ 
176 & 10:43:29.19 & $-$59:36:11.067 & $-$24.959 & 0.87 & 0.64 & 134 & 0.29 & 0.08 & 0.08 \\ 
177 & 10:43:27.84 & $-$59:35:42.698 & $-$24.280 & 0.60 & 0.50 & 178 & 0.14 & 0.08 & 0.02 \\ 
178 & 10:43:28.45 & $-$59:36:23.864 & $-$23.771 & 0.64 & 0.49 &  57 & 0.28 & 0.08 & 0.05 \\ 
179 & 10:43:30.24 & $-$59:36:04.331 & $-$24.055 & 2.33 & 0.87 & 171 & 0.28 & 0.08 & 0.22 \\ 
180 & 10:43:28.61 & $-$59:36:22.207 & $-$25.418 & 0.71 & 0.62 & 173 & 0.23 & 0.08 & 0.05 \\ 
181 & 10:43:30.37 & $-$59:35:33.977 & $-$24.440 & 0.76 & 0.53 &  69 & 0.34 & 0.08 & 0.07 \\ 
182 & 10:43:29.86 & $-$59:35:49.328 & $-$23.632 & 0.65 & 0.58 & 179 & 0.38 & 0.07 & 0.07 \\ 
183 & 10:43:30.04 & $-$59:36:02.124 & $-$24.422 & 0.92 & 0.47 & 166 & 0.31 & 0.08 & 0.07 \\ 
184 & 10:43:30.48 & $-$59:35:59.845 & $-$23.751 & 1.11 & 0.63 &   2 & 0.22 & 0.08 & 0.07 \\ 
185 & 10:43:30.83 & $-$59:35:22.508 & $-$23.597 & 1.71 & 0.68 & 151 & 0.15 & 0.08 & 0.09 \\ 
186 & 10:43:31.81 & $-$59:35:21.667 & $-$23.275 & 0.61 & 0.43 & 171 & 0.24 & 0.07 & 0.04 \\ 
187 & 10:43:32.00 & $-$59:36:00.095 & $-$25.106 & 1.80 & 0.98 & 100 & 0.32 & 0.08 & 0.28 \\ 
188 & 10:43:30.54 & $-$59:36:01.414 & $-$26.274 & 0.96 & 0.55 &  24 & 0.16 & 0.08 & 0.04 \\ 
189 & 10:43:30.45 & $-$59:35:28.837 & $-$24.929 & 0.78 & 0.55 &  25 & 0.35 & 0.08 & 0.06 \\ 
190 & 10:43:29.84 & $-$59:35:28.454 & $-$24.968 & 0.69 & 0.64 &   4 & 0.20 & 0.07 & 0.05 \\ 
191 & 10:43:30.00 & $-$59:35:34.039 & $-$25.609 & 0.72 & 0.65 &   7 & 0.36 & 0.07 & 0.07 \\ 
192 & 10:43:28.17 & $-$59:36:02.992 & $-$25.127 & 0.65 & 0.48 & 150 & 0.22 & 0.07 & 0.04 \\ 
193 & 10:43:30.41 & $-$59:35:30.020 & $-$25.932 & 1.19 & 0.42 & 107 & 0.21 & 0.07 & 0.06 \\ 
194 & 10:43:28.88 & $-$59:35:28.749 & $-$25.422 & 0.96 & 0.69 & 137 & 0.24 & 0.07 & 0.09 \\ 
\\
195 & 10:43:30.41 & $-$59:35:31.170 & $-$24.790 & 0.66 & 0.53 &  44 & 0.22 & 0.07 & 0.04 \\ 
196 & 10:43:30.26 & $-$59:35:56.592 & $-$23.387 & 1.02 & 0.63 &  21 & 0.24 & 0.07 & 0.08 \\ 
197 & 10:43:30.95 & $-$59:35:20.255 & $-$24.442 & 0.65 & 0.39 & 112 & 0.17 & 0.07 & 0.02 \\ 
198 & 10:43:30.10 & $-$59:35:31.372 & $-$23.463 & 0.73 & 0.67 & 179 & 0.28 & 0.07 & 0.06 \\ 
199 & 10:43:28.91 & $-$59:35:59.111 & $-$24.434 & 0.77 & 0.48 & 111 & 0.27 & 0.07 & 0.04 \\ 
200 & 10:43:29.65 & $-$59:36:19.842 & $-$25.444 & 1.24 & 0.84 &  88 & 0.27 & 0.07 & 0.14 \\ 
201 & 10:43:30.25 & $-$59:35:46.140 & $-$25.778 & 0.60 & 0.45 &  83 & 0.32 & 0.07 & 0.03 \\ 
202 & 10:43:29.08 & $-$59:35:59.030 & $-$23.786 & 1.04 & 0.61 &  70 & 0.26 & 0.06 & 0.08 \\ 
203 & 10:43:28.48 & $-$59:36:02.523 & $-$26.095 & 0.70 & 0.42 & 163 & 0.20 & 0.07 & 0.03 \\ 
204 & 10:43:29.20 & $-$59:35:45.345 & $-$24.596 & 0.77 & 0.44 & 147 & 0.17 & 0.07 & 0.03 \\ 
205 & 10:43:32.10 & $-$59:35:54.949 & $-$23.786 & 0.67 & 0.45 & 127 & 0.34 & 0.06 & 0.05 \\ 
206 & 10:43:30.53 & $-$59:36:03.120 & $-$23.420 & 0.81 & 0.63 &  26 & 0.31 & 0.07 & 0.07 \\ 
207 & 10:43:29.29 & $-$59:36:18.435 & $-$25.588 & 1.11 & 0.86 &  53 & 0.18 & 0.06 & 0.08 \\ 
208 & 10:43:28.11 & $-$59:35:56.623 & $-$23.102 & 0.88 & 0.62 & 108 & 0.30 & 0.06 & 0.08 \\ 
209 & 10:43:30.85 & $-$59:35:41.846 & $-$25.468 & 0.91 & 0.47 & 116 & 0.20 & 0.07 & 0.04 \\ 
210 & 10:43:29.63 & $-$59:35:29.354 & $-$24.446 & 0.62 & 0.49 &  50 & 0.25 & 0.06 & 0.03 \\ 
211 & 10:43:31.58 & $-$59:35:46.144 & $-$23.278 & 0.79 & 0.57 & 159 & 0.24 & 0.07 & 0.06 \\ 
212 & 10:43:30.45 & $-$59:36:17.447 & $-$25.765 & 0.79 & 0.39 & 130 & 0.22 & 0.06 & 0.03 \\ 
213 & 10:43:29.74 & $-$59:35:41.147 & $-$22.789 & 0.51 & 0.44 &   9 & 0.17 & 0.06 & 0.02 \\ 
214 & 10:43:28.30 & $-$59:36:22.158 & $-$23.282 & 0.78 & 0.56 &  26 & 0.37 & 0.06 & 0.07 \\ 
215 & 10:43:29.96 & $-$59:36:18.443 & $-$26.251 & 0.72 & 0.57 &  80 & 0.25 & 0.06 & 0.04 \\ 
216 & 10:43:30.51 & $-$59:35:55.600 & $-$23.116 & 0.70 & 0.47 & 163 & 0.34 & 0.06 & 0.05 \\ 
217 & 10:43:30.66 & $-$59:35:42.363 & $-$24.120 & 0.70 & 0.48 &  62 & 0.23 & 0.06 & 0.03 \\ 
218 & 10:43:29.26 & $-$59:35:59.906 & $-$25.437 & 0.74 & 0.42 & 131 & 0.31 & 0.06 & 0.04 \\ 
219 & 10:43:28.95 & $-$59:35:54.749 & $-$23.609 & 0.87 & 0.66 &  47 & 0.28 & 0.06 & 0.06 \\ 
220 & 10:43:32.08 & $-$59:35:51.953 & $-$24.112 & 0.69 & 0.39 &  50 & 0.37 & 0.06 & 0.04 \\ 
221 & 10:43:28.92 & $-$59:36:07.849 & $-$23.627 & 0.89 & 0.66 &  61 & 0.18 & 0.06 & 0.04 \\ 
222 & 10:43:29.09 & $-$59:36:16.425 & $-$25.111 & 0.78 & 0.62 & 113 & 0.27 & 0.06 & 0.05 \\ 
223 & 10:43:27.96 & $-$59:35:55.725 & $-$22.304 & 0.69 & 0.61 & 110 & 0.15 & 0.06 & 0.02 \\ 
224 & 10:43:29.13 & $-$59:35:38.040 & $-$25.110 & 0.68 & 0.42 &  24 & 0.17 & 0.06 & 0.02 \\ 
225 & 10:43:28.01 & $-$59:35:59.159 & $-$22.614 & 0.64 & 0.43 & 113 & 0.30 & 0.06 & 0.03 \\ 
226 & 10:43:30.68 & $-$59:35:53.313 & $-$24.278 & 0.66 & 0.38 &  42 & 0.65 & 0.06 & 0.03 \\ 
227 & 10:43:29.32 & $-$59:36:12.854 & $-$23.789 & 0.77 & 0.59 & 170 & 0.25 & 0.05 & 0.04 \\ 
228 & 10:43:29.83 & $-$59:35:59.485 & $-$23.616 & 1.09 & 0.64 &  10 & 0.29 & 0.06 & 0.08 \\ 
229 & 10:43:28.65 & $-$59:36:01.780 & $-$24.436 & 0.76 & 0.60 & 172 & 0.23 & 0.05 & 0.04 \\ 
230 & 10:43:28.98 & $-$59:36:14.733 & $-$24.443 & 0.79 & 0.47 & 135 & 0.23 & 0.05 & 0.04 \\ 
231 & 10:43:31.69 & $-$59:35:25.539 & $-$23.613 & 0.64 & 0.43 &  49 & 0.16 & 0.05 & 0.01 \\ 
232 & 10:43:30.16 & $-$59:36:14.880 & $-$25.420 & 0.99 & 0.59 &   8 & 0.24 & 0.05 & 0.06 \\ 
233 & 10:43:29.47 & $-$59:36:07.758 & $-$23.455 & 0.82 & 0.50 &  74 & 0.20 & 0.05 & 0.03 \\ 
234 & 10:43:29.68 & $-$59:35:37.741 & $-$24.119 & 0.88 & 0.48 &  42 & 0.23 & 0.05 & 0.03 \\ 
235 & 10:43:28.55 & $-$59:35:56.231 & $-$22.796 & 0.87 & 0.73 &   2 & 0.20 & 0.05 & 0.05 \\ 
236 & 10:43:32.08 & $-$59:35:56.719 & $-$24.116 & 0.68 & 0.56 &  34 & 0.29 & 0.05 & 0.04 \\ 
237 & 10:43:30.20 & $-$59:35:34.021 & $-$25.274 & 0.66 & 0.55 &  90 & 0.63 & 0.05 & 0.07 \\ 
238 & 10:43:29.88 & $-$59:36:15.442 & $-$24.788 & 0.70 & 0.55 & 169 & 0.25 & 0.05 & 0.04 \\ 
239 & 10:43:32.20 & $-$59:35:19.751 & $-$21.458 & 0.50 & 0.45 & 175 & 0.20 & 0.05 & 0.01 \\ 
240 & 10:43:30.35 & $-$59:35:28.017 & $-$25.759 & 0.61 & 0.44 &  75 & 0.28 & 0.05 & 0.03 \\ 
241 & 10:43:29.97 & $-$59:35:32.070 & $-$24.103 & 0.76 & 0.43 &  16 & 0.41 & 0.05 & 0.03 \\ 
242 & 10:43:28.84 & $-$59:35:57.535 & $-$23.288 & 0.86 & 0.60 & 114 & 0.21 & 0.05 & 0.04 \\ 
243 & 10:43:30.92 & $-$59:36:11.429 & $-$26.446 & 0.82 & 0.40 & 161 & 0.20 & 0.05 & 0.03 \\ 
244 & 10:43:30.28 & $-$59:36:10.551 & $-$24.120 & 0.71 & 0.40 & 119 & 0.19 & 0.05 & 0.02 \\ 
245 & 10:43:30.70 & $-$59:36:04.747 & $-$23.289 & 0.73 & 0.47 &  12 & 0.26 & 0.05 & 0.04 \\ 
\\
246 & 10:43:30.17 & $-$59:36:19.546 & $-$25.769 & 0.74 & 0.52 &  97 & 0.20 & 0.05 & 0.03 \\ 
247 & 10:43:29.16 & $-$59:35:37.835 & $-$24.104 & 0.72 & 0.52 &  29 & 0.21 & 0.05 & 0.03 \\ 
248 & 10:43:31.63 & $-$59:35:47.525 & $-$22.793 & 0.64 & 0.48 & 167 & 0.25 & 0.05 & 0.03 \\ 
249 & 10:43:31.25 & $-$59:36:03.136 & $-$24.114 & 0.65 & 0.48 &  82 & 0.47 & 0.05 & 0.03 \\ 
250 & 10:43:27.93 & $-$59:35:42.927 & $-$22.447 & 0.63 & 0.48 &  63 & 0.18 & 0.05 & 0.02 \\ 
251 & 10:43:28.21 & $-$59:36:02.033 & $-$25.607 & 0.78 & 0.52 &  94 & 0.20 & 0.05 & 0.03 \\ 
252 & 10:43:28.57 & $-$59:35:54.036 & $-$23.115 & 0.65 & 0.53 &  72 & 0.25 & 0.05 & 0.03 \\ 
253 & 10:43:32.31 & $-$59:35:19.252 & $-$23.117 & 0.76 & 0.37 & 111 & 0.23 & 0.05 & 0.01 \\ 
254 & 10:43:30.69 & $-$59:35:42.455 & $-$25.781 & 0.71 & 0.57 & 114 & 0.25 & 0.05 & 0.03 \\ 
\enddata
\end{deluxetable*}

\appendix

\section{Summary of the observations}

Table~\ref{tab:obs} lists the main properties of ALMA interferometric observations
reported in this paper, including the frequency bands (1), the antenna
array (2), the observation dates (3), the complex gain calibrator (4), bandpass calibrators (5), and
the maximum (6) and minimum (7) antenna baselines.

\begin{deluxetable*}{lcccccc}[h!]
\tablecaption{Summary of the Interferometric Observations \label{tab:obs}}
\tablehead{
\colhead{Band\vspace {-0.1in}} & \colhead{Array} &\colhead{Date} & \colhead{Gain cal.} & \colhead{Bandpass cal.} & \colhead{Max. Baseline} & \colhead{Min. Baseline}\\
\colhead{(1)} & \colhead{(2)} &\colhead{(3)} & \colhead{(4)} & \colhead{(5)} & \colhead{(6)} & \colhead{(7)}
}
\startdata
\noalign{\smallskip}
6 & 12-m & 19-Dec-2015 & J1047-6217 & J1107-4449 & 6300m & 15.1m\\
 & & 02-Mar-2016 & J1047-6217 & J1107-4449 & 390.0m & 15.2m\\
 & & 17-Mar-2016 & J1047-6217 & J1107-4449 & 460.0m & 15.2m\\
\noalign{\smallskip}
\tableline
\noalign{\smallskip}
6 & 7-m & 28-Jun-2016 & J1147-6753 & J0854+2006 & 49.0m & 8.8m\\
 & & 30-Jun-2016 & J1147-6753 & J0854+2006 & 43.2m & 8.8m\\
 & & 30-Jun-2016 & J1147-6753 & J1058+0133 & 43.2m & 8.8m\\
 & & 30-Jun-2016 & J1147-6753 & J1427-4206 & 43.2m & 8.8m\\
\noalign{\smallskip}
\tableline
\noalign{\smallskip}
8 & 12-m & 19-Dec-2015 & J0490-6107 & J1256-0547 & 6300m & 15.1m\\
\noalign{\smallskip}
\tableline
\noalign{\smallskip}
8 & 7-m & 09-Aug-2016 & J1047-6217 & J0522-3627 & 45.0m & 8.9m\\
 & & 20-Aug-2016 & J1047-6217 & J1256-0547 & 45.0m & 8.9m\\
 & & 21-Aug-2016 & J1047-6217 & J1256-0547 & 45.0m & 8.9m\\
 & & 25-Aug-2016 & J1047-6217 & J1256-0547 & 45.0m & 8.9m\\
 & & 07-Sep-2016 & J1047-6217 & J0522-3627 & 45.0m & 8.9m\\
 & & 08-Sep-2016 & J1047-6217 & J1256-0547 & 45.0m & 8.9m\\
\noalign{\smallskip}
\enddata
\end{deluxetable*}

\section{Combining single-dish and interferometric data}

In this section we describe the procedure followed to combine single-dish and
interferometric ALMA data.  As a preliminary step, we independently imaged
12-m and 7-m array observations of the targeted emission lines. This step is not
necessary for combining the data but it provides images of the high spatial
frequency components of the emission that can be compared to the combined
images. All the images discussed in this section were produced using the task
TCLEAN provided in version 5.6.0 of the CASA package.

The best practice for combining single-dish and interferometric data is matter
of debate. Several techniques have
been proposed, including combining data before, during, or after deconvolution
\citep[see][for a recent review]{tp2vis}.
In this paper, we combined interferometric and single-dish data in the
Fourier space before deconvolution using the publicly available
{\it Total Power to Visibilities} (TP2VIS) algorithm presented in \cite{tp2vis}.
This method requires deconvolving the single-dish image with the single-dish beam
to obtain an infinite resolution model for the emission. This model is then 
Fourier-transformed to obtain a set of ``single-dish'' visibilities which are concatenated
to the interferometric visibilities and imaged using a standard CLEAN algorithms.

The publicly available version of TP2VIS performs the deconvolution of single-dish
images by Fourier-transforming them and dividing the resulting visibilities by
the Fourier transform of the synthesized beam. We find that this direct approach to deconvolution
does not work well with our data because the line emission extends to the edge of the image field
in most of the channels.  In such cases, the Fourier transform of a
single-dish image acquires strong high frequency components owing to the sharp cutoff of the
emission at the image edges. TP2VIS provides a few tapering options to mitigate
this issue but none of them led to acceptable results for this data set.

Instead, we deconvolved the single-dish data using an algorithm
similar to CLEAN. We performed the decolvolution in the image plane by
subtracting point source components. The model image is then Fourier transformed
using TP2VIS. The resulting visibilties are combined to the interferometric data
and imaged using TCLEAN. We checked the fidelity of the single-dish image deconvolution
by re-imaging the calculated single-dish visibilities using TCLEAN and
comparing the resulting images to the original single-dish maps. In general, we
find that the final images differ from the original images only by a few percent.

\section{Optical Depth Calculations}

This section calculates the optical depth at line center for a few transitions
of interest in PDRs and molecular clouds, assuming standard solar abundances \citep{abund}
and a thermal broadening profile for a H-column density of $2.7\times 10^{21}$ cm$^{-2}$,
which corresponds to a visual extinction A$_V$ = 1 \citep{Liszt14}.
The derivation largely follows that of Chapter~10 of \citet{rybicki}. 

The optical depth $\tau_\nu$ at frequency $\nu$ is related to the differential
path length $dl$ and the opacity $\alpha_\nu$ via $d\tau_\nu$ = $\alpha_\nu dl$,
so for a uniform gas with constant path length $L$,
the optical depth integrated along the line of sight becomes

\begin{equation}
\label{eq:tau}
\tau_\nu = \alpha_\nu L
\end{equation}

\noindent 
The opacity is related to the Einstein B
coefficients between the upper level (denoted by 2) and lower level
(denoted by 1), and their corresponding number densities $n_1$ and $n_2$ via

\begin{equation}
\label{eq:alpha}
\alpha_\nu = {\frac{h\nu}{4\pi}\left[n_1B_{12}-n_2B_{21}\right]\phi(\nu)}
\end{equation}

\noindent
where the normalized line profile $\phi(\nu)$ for thermal broadening
is 

\begin{equation}
\label{eq:phi}
\phi(\nu) = \frac{1}{\Delta\nu_D \sqrt{\pi}}e^{-\left[(\nu-\nu_\circ)^2/(\Delta\nu_D)^2\right]}
\end{equation}

\noindent
and 

\begin{equation}
\label{eq:deltanud}
\Delta\nu_D = \frac{\nu_\circ}{c}\sqrt{\frac{2kT}{m}}
\end{equation}

\noindent
is the Doppler broadening coefficient. Here,
$\nu_\circ$ is the frequency of the transition at line center,
m is the mass of the atom or molecule, and k is Boltzman's constant and
T is the temperature.

Using the Einstein relations

\begin{equation}
\label{eq:Einstein}
A_{21} = \frac{2h\nu^3}{c^2}B_{21}\ ;\ B_{21} = \frac{g_1}{g_2}B_{12}
\end{equation}

\noindent
we can rewrite the term inside the brackets in Equation~\ref{eq:alpha} as

\begin{equation}
\label{eq:squiggle}
n_1B_{12}-n_2B_{21} = \frac{g_2n_1A_{21}c^2}{2g_1h\nu^3}{\left[1 - \eta e^{-{h\nu/kT}}\right]}
\end{equation}

\noindent
where we define

\begin{equation}
\label{eq:eta}
\eta = \left(\frac{n_2}{n_1}\right)\bigg/\left(\frac{n_2}{n_1}\right)_{LTE}
\end{equation}

\noindent
and

\begin{equation}
\label{eq:lte}
\left(\frac{n_2}{n_1}\right)_{LTE} = \ \ \frac{g_2}{g_1}{\rm e}^{-{h\nu/kT}}
\end{equation}

\noindent
Defining

\begin{equation}
\label{eq:S}
S  = 1 - \eta e^{-{h\nu/kT}}
\end{equation}

\noindent 
and combining equations \ref{eq:tau}, \ref{eq:alpha}, \ref{eq:phi}, \ref{eq:deltanud},
\ref{eq:squiggle} and \ref{eq:S} leads to

\begin{equation}
\label{eq:tauanalytic}
\tau_\circ = \frac{\lambda^3A_{21}g_2}{8\pi g_1}\left(\frac{m}{2\pi kT}\right)^{\frac{1}{2}}S{\rm N}_1
\end{equation}
\noindent 
for the optical depth at line center, where the column density in the lower level
is N$_1$ = $n_1L$. Normalizing N$_1$ by the total column density N$_{{\rm TOT}}$
of the atom or molecule, and noting that the ratio 
N$_{{\rm TOT}}$/N$_{\rm H}$ is the abundance of the specie relative to hydrogen,
when we insert numbers into Equation~\ref{eq:tauanalytic} we obtain

\begin{multline}
\label{eq:taufinal}
\tau_\circ = 1741
\left[\frac{\lambda}{1\ \mu {\rm m}}\right]^3
\left[\frac{A_{21}}{1\ \rm{sec}^{-1}}\right]
\left[\frac{g_2}{g_1}\right]
\left[\frac{m}{1\ {\rm amu}}\right]^{1/2}
\\
\times\
\left[\frac{T}{1\ {\rm K}}\right]^{-1/2}
\left[\frac{S}{1}\right]
\left[\frac{{\rm N}_1}{{\rm N}_{\rm TOT}}\right]
\left[\frac{{\rm N}_{\rm TOT}}{{\rm N}_{\rm H}}\right]
\left[\frac{{\rm N}_{\rm H}}{10^{21} {\rm cm}^{-2}}\right]
\end{multline}

\begin{deluxetable*}{cccccccccc}[ht]
\tablecaption{Optical Depth Calculations for A$_V$ = 1 at Line Center\label{tab:tau}}
\tablehead{
\colhead{Specie}\vspace {-0.08in} & \colhead{$\lambda$ ($\mu$m)} & \colhead{A$_{21}$ (s$^{-1}$)}
 & \colhead{g2/g1} & \colhead{m (amu)} & \colhead{T (K)}
 & \colhead{S} & \colhead{N$_1$/N$_{TOT}$}
 & \colhead{N$_{TOT}$/N$_H$} & \colhead{$\tau_\circ$}\\
 (1)\vspace {-0.1 in}&(2)&(3)&(4)&(5)&(6)&(7)&(8)&(9)&(10)\\
}
\startdata
\noalign{\smallskip}
 $^{12}$CO (2-1)& 1300 & $6.91\times 10^{-7}$& 1.67&28 & 30   & 0.308& 0.22& $1.34\times 10^{-4}$ & 108 \\
{[C II]} & 158        & $2.29\times 10^{-6}$& 2   &12 & 100   & 0.598& 0.55& $2.69\times 10^{-4}$ & 2.6 \\
$^{13}$CO (2-1)& 1360 & $6.04\times 10^{-7}$& 1.67&29 & 30    & 0.297& 0.22& $1.76\times 10^{-6}$ & 1.3 \\
{[C I]} & 609         & $7.93\times 10^{-8}$& 3   &12 & 30    & 0.545& 0.33& $2.69\times 10^{-5}$ & 0.77\\
C$^{18}$O (2-1)& 1366 & $6.01\times 10^{-7}$& 1.67&30 & 30    & 0.296& 0.22& $2.44\times 10^{-7}$ & 0.19 \\
{[O I]} & 63.1        & $8.91\times 10^{-5}$& 0.6 &16 & $10^4$& 0.023& 0.53& $4.90\times 10^{-4}$ & 0.015 \\
\textquotedbl & \textquotedbl & \textquotedbl & \textquotedbl & \textquotedbl & 100 & 0.898 & 0.93 & \textquotedbl & 10.3 \\
{[S II]} & 0.673      & $7.84\times 10^{-4}$&   1 &32 & $10^4$& 0.882& 0.75 & $1.62\times 10^{-5}$ & $6.8\times 10^{-7}$ \\
\noalign{\smallskip}
\enddata
\end{deluxetable*}


We present the results of Equation~\ref{eq:taufinal} in Table~\ref{tab:tau} for
the lines observed in this paper, for PDR lines [O~I] 63~$\mu$m and [C~II] 158~$\mu$m,
and for [S~II] 6731\AA , a typical optical forbidden line transition.
The A-values and wavelengths were taken from the 2020 version of the LAMDA database
\citep{lamda} and from \citet{mendoza83}. The temperatures in the Table
represent typical values for molecular clouds (for CO and C~I), PDRs (for C~II and O~I),
and H~II regions or radiative shocks (for O~I and S~II).
The partition function needed to find N$_1$/N$_{{\rm TOT}}$
assumes the transition is in its high-density limit, so a Boltzmann distribution describes
the level populations (i.e., $\eta$ = 1 in Equation~\ref{eq:eta}).
If this condition is not satisfied, the optical depths at line center in
column 10 will be lower than those tabulated here.  

We take all C to be C~II, all S to be S~II, and all O to be O~I for calculations of
[C~II] 158~$\mu$m, [S~II] 0.673~$\mu$m and [O~I] 63~$\mu$m, respectively.
As described in Sec.~\ref{sec:ww}, for CO and C~I within the molecular cloud
we adopt an abundance ratio of H$_2$:CO = 3700 \citep{lacy94},
which for modern solar elemental abundances \citep{abund} means that 50\%\ of carbon resides in CO.
We then take C~I:CO $\sim$ 0.2, so 10\%\ of carbon is in C~I, with the remaining 40\%\ of
carbon locked in grains. The actual ratios could vary substantially between different regions.
Finally, for the CO isotopolgues
we adopt abundance ratios of $^{12}$CO/$^{13}$CO = 77 and $^{12}$CO/C$^{18}$O = 550.

The table shows we expect $^{12}$CO to be very optically thick, $^{13}$CO to be optically
thick in some places and thin in others, [C~I] to have optical depth of order unity, and
C$^{18}$O to be optically thin. 
Typical optical emission lines such as [S~II] 6731\AA\ are very optically thin, while
fine structure atomic lines such as [O~I] 63~$\mu$m and [C~II] 158~$\mu$m may be optically
thick or thin depending upon how broadened the lines are in the emitting region.

\software{TCLEAN(CASA version 5.6.0), TP2VIS (Koda et~al. 2019), 
Starlink (Currie et~al. 2014), CUPID (Berry et~al. 2007)}

\acknowledgments

%

We would like to thank David Berry for the help provided in using the CUPID
package and Jin Koda for his help is using the tp2vis package. A.~I. acknowledges
support from the ALMA Study Project \#358232. A.~I. and M.~H. acknowledge support
from the National Science Foundation under grant No. AST-1715719. This paper
makes use of the following ALMA data: ADS/JAO.ALMA\#2015.1.00656.S. ALMA is a
partnership of ESO (representing its member states), NSF (USA) and NINS
(Japan), together with NRC (Canada), MOST and ASIAA (Taiwan), and KASI
(Republic of Korea), in cooperation with the Republic of Chile. The Joint ALMA
Observatory is operated by ESO, AUI/NRAO and NAOJ. The National Radio Astronomy
Observatory is a facility of the National Science Foundation operated under
cooperative agreement by Associated Universities, Inc. The authors acknowledge
the data analysis facilities provided by the Starlink Project which is run by
CCLRC on behalf of PPARC. In addition, the following Starlink packages have
been used: CUPID/GAUSSCLUMP.

\vspace{5mm}
\facilities{ALMA}

\end{document}